\documentclass[aps,showpacs,prx,superscriptaddress,preprintnumbers,amsmath,amssymb,twocolumn,nofootinbib]{revtex4-2}
\usepackage[utf8]{inputenc}
\usepackage[T1]{fontenc}
\usepackage{graphicx}
\usepackage{dcolumn}
\usepackage{appendix}
\usepackage{amsmath}
\usepackage{bm}
\usepackage[dvipsnames]{xcolor}

\newcommand{\Rbar}{\lvert\bar{R}\rangle} 
\newcommand{\Lbar}{\lvert\bar{L}\rangle}

\usepackage{subfigure}
\usepackage{bbm}
\usepackage{enumerate}

\usepackage[
bookmarks=true,
colorlinks,
linkcolor=black,
urlcolor=black,
citecolor=black,
plainpages=false,
pdfpagelabels,
final,
breaklinks=true
]{hyperref}

\usepackage{titlesec}

\usepackage{array}

\usepackage{physics}

\begin{document}
	\allowdisplaybreaks
	
	\title{Quantum optical analysis of high-order harmonic generation in H$_{\textbf{2}}^{\vb{+}}$ molecular ions}
	
	\author{J.~Rivera-Dean}
	\email{javier.rivera@icfo.eu}
	\affiliation{ICFO -- Institut de Ciencies Fotoniques, The Barcelona Institute of Science and Technology, 08860 Castelldefels (Barcelona)}
	
	\author{P.~Stammer}
	\affiliation{ICFO -- Institut de Ciencies Fotoniques, The Barcelona Institute of Science and Technology, 08860 Castelldefels (Barcelona)}
	
	\author{A.~S.~Maxwell}
	\affiliation{Department of Physics and Astronomy, Aarhus University, DK-8000 Aarhus C, Denmark}
	
	\author{Th.~Lamprou}
	\affiliation{Foundation for Research and Technology-Hellas, Institute of Electronic Structure \& Laser, GR-70013 Heraklion (Crete), Greece}
	\affiliation{Department of Physics, University of Crete, P.O. Box 2208, GR-70013 Heraklion (Crete), Greece}
	
	\author{E.~Pisanty}
	\affiliation{Attosecond Quantum Physics Laboratory, Department of Physics, King's College London, Strand WC2R 2LS, London, United Kingdom}
	
	\author{P.~Tzallas}
	\affiliation{Foundation for Research and Technology-Hellas, Institute of Electronic Structure \& Laser, GR-70013 Heraklion (Crete), Greece}
	\affiliation{ELI-ALPS, ELI-Hu Non-Profit Ltd., Dugonics tér 13, H-6720 Szeged, Hungary}
	
	\author{M.~Lewenstein}
	\email{maciej.lewenstein@icfo.eu}
	\affiliation{ICFO -- Institut de Ciencies Fotoniques, The Barcelona Institute of Science and Technology, 08860 Castelldefels (Barcelona)}
	\affiliation{ICREA, Pg. Llu\'{\i}s Companys 23, 08010 Barcelona, Spain}
	
	\author{M.~F.~Ciappina}
	\email{marcelo.ciappina@gtiit.edu.cn}
	\affiliation{Physics Program, Guangdong Technion--Israel Institute of Technology, Shantou, Guangdong 515063, China}
	\affiliation{Technion -- Israel Institute of Technology, Haifa, 32000, Israel}
	\affiliation{Guangdong Provincial Key Laboratory of Materials and Technologies for Energy Conversion, Guangdong Technion – Israel Institute of Technology, Shantou, Guangdong 515063, China}

	\date{\today}
	\begin{abstract}
        We present a comprehensive theoretical investigation of high-order harmonic generation in H$_2^+$ molecular ions within a quantum optical framework. Our study focuses on characterizing various quantum optical and quantum information measures, highlighting the versatility of HHG in two-center molecules towards quantum technology applications. We demonstrate the emergence of entanglement between electron and light states after the laser-matter interaction. We also identify the possibility of obtaining non-classical states of light in targeted frequency modes by conditioning on specific electronic quantum states, which turn out to be crucial in the generation of highly non-classical entangled states between distinct sets of harmonic modes. Our findings open up avenues for studying strong-laser field-driven interactions in molecular systems, and suggest their applicability to quantum technology applications.
	\end{abstract}
	\maketitle
	
\section{INTRODUCTION}
High-order harmonic generation (HHG) arises from the highly nonlinear interaction between an intense, short laser pulse and a target, which can be either a gaseous system composed of atoms or molecules, as well as solid-state systems and nanostructures \cite{lhuillier_theoretical_1991,lynga_high-order_1996,krausz_attosecond_2009,ghimire_observation_2011,vampa_high-harmonic_2015,ciappina_attosecond_2017,amini_symphony_2019}. Currently, HHG serves as one of the main methods for generating spatially and temporally coherent extreme-ultraviolet (XUV) light, as well as subfemtosecond and attosecond pulses \cite{corkum_attosecond_2007}. Coherent light sources spanning the ultraviolet (UV) to XUV spectral range, find wide applications in various fields, including fundamental research, material science, biology, and lithography \cite{krausz_attosecond_2009}. 

The fundamental physics underlying the HHG process is commonly described by the \emph{three-step model} or \emph{simple man's model} \cite{krause_high-order_1992,corkum_plasma_1993,Kulander1993}. According to this model, we have that when an atom or molecule interacts with a strong laser pulse, an electron is liberated through tunnel ionization, typically during the peak of the laser's electric field within an optical cycle. The freed electron is then driven away from the ionic core accelerated by the laser field, following an oscillating trajectory. Along this trajectory, the electron gains kinetic energy, which is subsequently released as high-energy radiation during the recombination process. Due to the periodic nature of the laser field, this three-step process repeats every half-cycle. The HHG process in atomic systems, has been extensively studied, and various theoretical models such as the strong-field approximation (SFA) \cite{lhuillier_high-order_1993,lewenstein_theory_1994,olga_simpleman,amini_symphony_2019}, have been developed to describe it, complementing the computationally expensive solution of the time-dependent Schrödinger equation (TDSE). In the case of molecular systems consisting of two-atoms, they have been extensively studied in the past either by using fully numerical methods \cite{ivanov_generation_1993,zuo_harmonic_1993,yu_threedimensional_1995,moreno_ultrahigh_1997,kopold_model_1998,alon_selection_1998,bandrauk_high-order_1999,lappas_orientation_2000,averbukh_high-order_2001,kreibich_even-harmonic_2001}, or by SFA extensions to the molecule scenario \cite{kjeldsen_strong-field_2004,milosevic_strong-field_2006,chirila_strong-field_2006,lein_molecular_2007,suarez_above-threshold_2016,suarez_high-order-harmonic_2017,suarez_above-threshold_2018,suarez_rojas_strong-field_2018,labeye_dynamical_2019}. These approaches have contributed to our understanding of HHG in molecular systems and provided valuable insights into their complex dynamics.

The interest in HHG processes in molecular targets, compared to their atomic counterpart, stems from the additional degrees of freedom it provides. For instance, molecular HHG involves the alignment of the molecular axis in relation to the polarization of the laser field, as well as the inherent multi-center nature of the strong-field process. On top of this, molecular HHG encodes valuable information about the electronic orbital structure, offering reliable means of extracting molecular intrinsic parameters with sub-Angstrom spatial and attosecond temporal resolutions \cite{lein_interference_2002,kanai_quantum_2005,lein_molecular_2007,torres_probing_2007,morishita_accurate_2008}. In this regard, it has been shown that the unique properties of molecular HHG spectra can be harnessed to extract structural information from simple molecules \cite{itatani_tomographic_2004}, while HHG spectroscopy has also shown potential for extracting structural and dynamical information from more complex targets \cite{ciappina_multislit_2007,laarmann_control_2007,ciappina_high-order_2008}. Lastly, studies on small molecules have successfully recovered the temporal evolution of electronic wave functions directly \cite{smirnova_high_2009,haessler_attosecond_2010,kraus_measurement_2015}.

Most of the methods developed to study the HHG processes in molecular systems consider a semiclassical approach, treating the target quantum mechanically, while considering the electromagnetic field a classical quantity. However, in recent years, there has been a growing interest in the quantum optical characterization of strong-field processes, revealing intriguing features such as the generation of non-classical quantum states of light \cite{lewenstein_generation_2021,rivera-dean_new_2021,rivera-dean_light-matter_2022,stammer_quantum_2023,pizzi_light_2022} with intensities high enough to drive nonlinear processes in matter~\cite{lamprou_nonlinear_2023}, hybrid entangled states between light and matter \cite{rivera-dean_light-matter_2022}, highly frequency-entangled states of light \cite{stammer_high_2022,stammer_theory_2022} and the influence of the photon statistics of the input driving field on the HHG spectrum and the associated electronic trajectories \cite{even_tzur_photon-statistics_2023}. These significant research efforts have underscored the potential of strong-field physics in atoms towards photonic-based quantum information science applications \cite{gilchrist_schrodinger_2004,gisin_quantum_2007,obrien_photonic_2009,lewenstein_attosecond_2022,bhattacharya_strong_2023}. Moreover, recent theoretical works have demonstrated that similar phenomena can be observed in solid-state materials as well \cite{gonoskov_nonclassical_2022,rivera-dean_entanglement_2023}. Specifically, in Ref.~\cite{rivera-dean_entanglement_2023} the delocalized nature of the recombination process in solid-state targets was shown to impact the final state of the field, potentially resulting in the generation of non-classical states of light and electron-light entangled states.

In this study, we investigate the extent to which these effects can be observed when utilizing symmetric diatomic molecules as targets of intense laser fields, where the active electron is now delocalized between the two centers. This scenario offers a simpler setup compared to the solid-state system, where the electron can recombine, to some extent \cite{brown_real-space_2022}, anywhere in the solid. Nevertheless, as we will see in the remainder of this work, HHG in symmetric diatomic molecules such as H$_2^+$ lead to interesting non-classical characteristics on the electromagnetic field modes that depend, in certain cases, on the final state of the electron. With this aim, we first characterize the interaction between the molecular system and the quantized field. Subsequently, we demonstrate how the final state of the electron influences the generation of non-classical states of light and the entanglement features in the post-interaction state. As we will see, these effects strongly rely on molecular features, such as the distance between the atomic centers and the number of molecules interacting with the field.

The paper is organized as follows. After this general introduction, we discuss the theoretical background in Section~\ref{Sec:Theory}, where we present a simplified, discrete mode description of the quantized electromagnetic field, and the relevant molecular states. Section~\ref{Sec:Results}, summarizes the main results of the paper: mean photon number in the single and many molecules regime, Wigner function distributions of different field modes, electron-light entanglement, and entanglement between different sets of frequency modes. We conclude in Section~\ref{Sec:Conclusions}. The paper contains two Appendices, \ref{App:First} and \ref{App:Final}, with more technical explanations. 

\section{THEORETICAL BACKGROUND}\label{Sec:Theory}

In this work, we consider the case where a diatomic molecule interacts with a strong-laser field with a peak intensity in the order of $10^{14}$ W/cm$^2$, and whose wavelength belongs to the near-infrared regime ($\lambda_L \sim 750 - 1400$ nm). The Hamiltonian characterizing the interaction, under the single-active electron (SAE), Born-Oppenheimer \cite{born_zur_1927,atkins_book} and dipole approximations is
\begin{equation}
    \hat{H}(t)
        = \hat{H}_{\text{mol}}
            + \hat{H}_{\text{int}}(t)
            + \hat{H}_{\text{field}},
\end{equation}
where $\hat{H}_{\text{mol}} = \hbar^2\hat{\boldsymbol{P}}^2/(2m) + V(\hat{\boldsymbol{R}})$ is the molecular Hamiltonian, with $m$ the electron's mass, $\hat{\boldsymbol{P}}$ the electronic momentum operator and $V(\hat{\boldsymbol{R}})$ the molecular potential; $H_\text{int}(t) = e \hat{\boldsymbol{R}}\cdot \hat{\boldsymbol{E}}(t)$ is the interaction Hamiltonian in the length gauge, with $e$ the electron's charge, and $\hat{H}_{\text{field}}$ is the electromagnetic free-field Hamiltonian. Here, we aim to describe interactions with laser pulses of finite duration, which ultimately requires the introduction of the full continuum spectrum of the electromagnetic field. However, for the sake of simplicity, we consider a discrete set of modes spanning from the central frequency of the driving laser $\omega_L$, up to the cutoff region of the harmonic spectrum, $\omega_{q_c} = q_c \omega_L$, i.e., $\{\omega_q = q\omega_L: q = 1,2,\cdots,q_c\}$. Thus, we write the free-field Hamiltonian for linearly polarized fields as $\hat{H}_{\text{field}} = \sum_{q} \hbar \omega_{q} \hat{a}^\dagger_{q}\hat{a}_{q}$, with $\hat{a}_{q}$ ($\hat{a}^\dagger_{q}$) the annihilation (creation) operator acting on the field mode with frequency $\omega_q$. In order to account for the pulse envelope of our driving field, we model the laser electric field operator as
\begin{equation}
    \hat{\boldsymbol{E}}(t)
        = -i f(t)\sum_{q}
             \vb{g}(\omega_q)
             (\hat{a}^\dagger_{q} e^{i\omega_q t}
             - \hat{a}_{q} e^{-i\omega_q t}),
\end{equation}
where $\vb{g}(\omega_q) \equiv \boldsymbol{\varepsilon}_{\mu}\sqrt{\hbar\omega_q/(2\epsilon_0 V)}$ is a factor arising from the expansion of the electric field operator into the field modes \cite{ScullyBook,Gerry__Book_2001}, with $\boldsymbol{\varepsilon}_{\mu}$ a unitary vector pointing in the direction along which the field is polarized, $\epsilon_0$ the vacuum permittivity and $V$ the quantization volume. Here, $0 \leq f(t) \leq 1$ is a dimensionless function describing the laser pulse envelope.
	
\begin{figure}
    \centering
    \includegraphics[width=1\columnwidth]{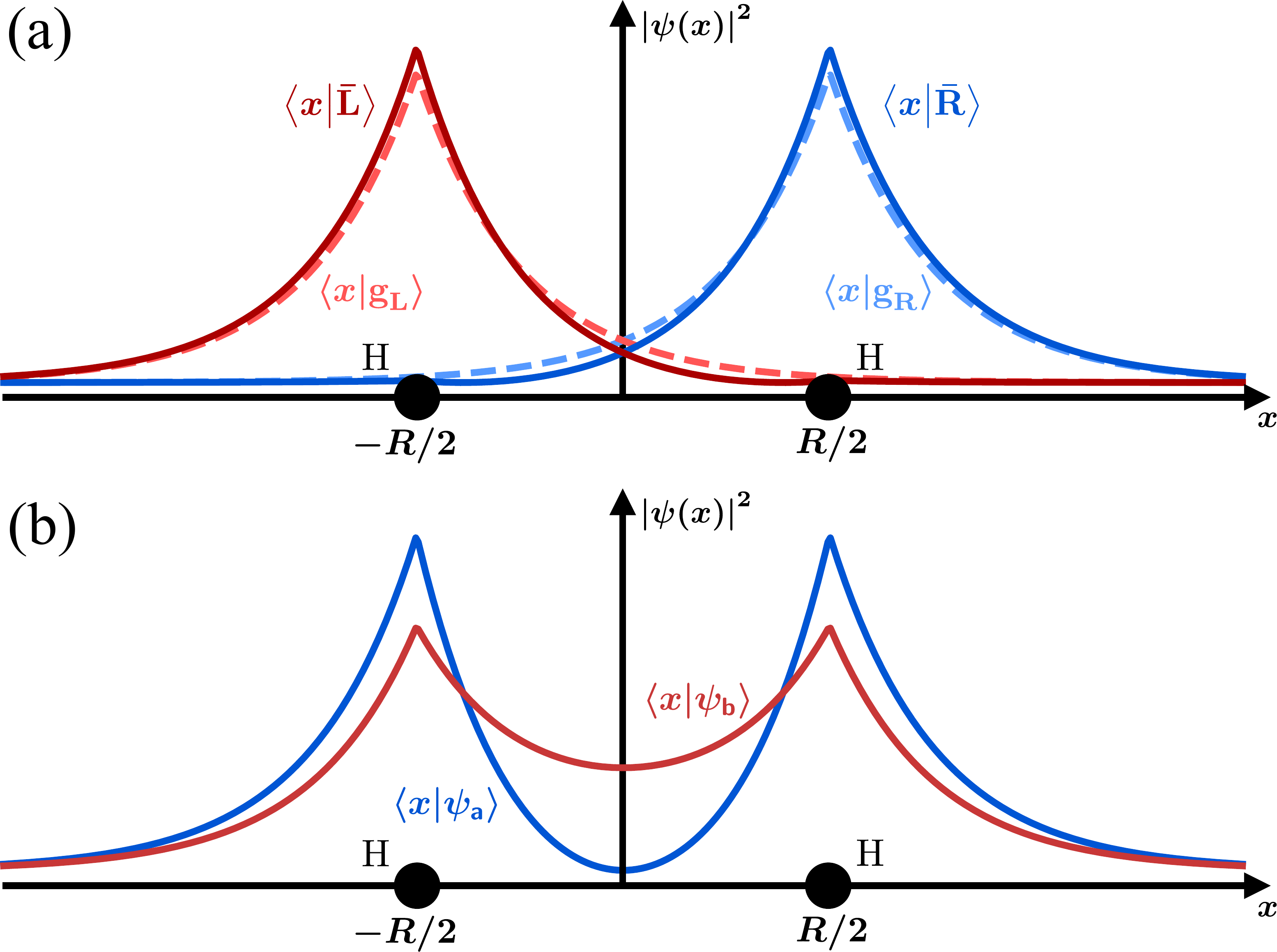}
    \caption{Schematic representation of a diatomic molecule. The two centers (H in the figure) are located at a distance $R$ away one from the other. In (a), the solid curves represent the spatial distribution of the atomic ground states $\{\ket{\text{g}_L}, \ket{\text{g}_R}\}$, while the dashed curves show the spatial distribution of the slightly more localized states $\{\lvert \bar{L}\rangle, \lvert \bar{R}\rangle\}$. These states satisfy orthonormality conditions, and are given as linear combinations of atomic ground state orbitals. In (b), the solid curves show the spatial distribution of bonding (red curve) and antibonding (blue curve) states, computed under the LCAO.}
    \label{Fig:Scheme}
\end{figure}

Within this framework, we describe the initial state of the electromagnetic field as $(\ket{\alpha} \bigotimes_{q=2}^{q_c}\ket{0_{q}})$, i.e., the fundamental IR mode is in a coherent state of amplitude $\alpha$, while the harmonic modes are unpopulated, i.e.~they are in a vacuum state. On the other hand, we set the molecule to initially be in its ground state. Here, we consider the case of the $\text{H}_2^+$ molecular ion. Its ground state, under the Linear Combination of Atomic Orbitals (LCAO) \cite{atkins_book,finkelstein_uber_1928}, is given by the so-called bonding state $\ket{\psi_{\mathsf{b}}} \propto \ket{\text{g}_R} + \ket{\text{g}_L}$, pictorially represented in Fig.~\ref{Fig:Scheme}~(b) with the red curve. This state  is given as the symmetric superposition of the ground state orbitals of each of the centers composing the molecule, namely \emph{right} ($\ket{\text{g}_R}$) and \emph{left} ($\ket{\text{g}_L}$) centers, represented in Fig.~\ref{Fig:Scheme}~(a) with the dashed curves. Alternatively, in terms of the LCAO, the first excited state of the molecule corresponds to the antisymmetric superposition of these ground state orbitals, that is, $\ket{\psi_{\mathsf{a}}} \propto \ket{\text{g}_R} - \ket{\text{g}_L}$ represented with the blue solid curve in Fig.~\ref{Fig:Scheme}~(b), and which we do take into account in our calculations. With all this, we write the joint initial state as
\begin{equation}\label{Eq:Init:cond}
    \ket{\Psi(t_0)}
        = \ket{\psi_{\mathsf{b}}}
            \otimes
                \Big[
                    \ket{\alpha}
                    \bigotimes^{q_c}_{q=2}
                        \ket{0_{q}}
                \Big].
\end{equation}

Within a more convenient frame, under which the Hamiltonian is given by
\begin{equation}
    \hat{H}(t)
        = \hat{H}_{\text{mol}} 
            + e \hat{\boldsymbol{R}}\cdot
             \big(
                \hat{\boldsymbol{E}}(t)
                + \boldsymbol{E}_{\text{cl}}(t)
             \big),
\end{equation}
with $\boldsymbol{E}_{\text{cl}}(t) = [\bra{\alpha}\bigotimes_{q=2}^{q_c}\bra{0_q}]\hat{\boldsymbol{E}}(t)[\ket{\alpha}\bigotimes_{q=2}^{q_c}\ket{0_q}]$, the initial state of the system can be rewritten as
\begin{equation}
\ket{\bar{\Psi}(t_0)}
    =\ket{\psi_{\mathsf{b}}}
        \bigotimes_{q}
            \ket{0_{q}},
\end{equation}
and the time-dependent Schrödinger equation reads
\begin{equation}\label{Eq:Sch:before}
    i\hbar \pdv{\ket{\bar{\Psi}(t)}}{t}
        = \Big[
                \hat{H}_{\text{mol}} 
                + e \hat{\boldsymbol{R}}\cdot
                \big(
                    \hat{\boldsymbol{E}}(t)
                    + \boldsymbol{E}_{\text{cl}}(t)
                \big)
            \Big]
                \ket{\bar{\Psi}(t)}.
	\end{equation}

In order to solve this differential equation, we move to the interaction picture with respect to the \emph{semiclassical} Hamiltonian $\hat{H}_{\text{sc}}(t) =  \hat{H}_{\text{mol}} + e \hat{\boldsymbol{R}}\cdot \boldsymbol{E}_{\text{cl}}(t)$, such that the position operator acquires a time dependence, i.e., $\hat{\boldsymbol{R}}(t) \equiv \hat{U}^\dagger_{\text{sc}}(t) \hat{\boldsymbol{R}} \hat{U}_{\text{sc}}(t)$ with $\hat{U}_{\text{sc}}(t) = \hat{\mathcal{T}} e^{-\frac{i}{\hbar}\int^t_{t_0} \dd \tau \hat{H}_{\text{sc}}(\tau)}$, where $\hat{\mathcal{T}}$ is the time-ordering operator. Similarly to Refs.~\cite{lewenstein_generation_2021,rivera-dean_strong_2022,stammer_theory_2022,stammer_quantum_2023} where a quantum optical characterization of atomic-HHG processes is done, we neglect the continuum population at all times. Here, we assume this contribution to be small in comparison to that of the molecular lowest energy states \cite{lewenstein_theory_1994,suarez_high-order-harmonic_2017,suarez_rojas_strong-field_2018}. Therefore, by projecting the Schrödinger equation obtained under this assumption with respect to the $\ket{\psi_{\mathsf{b}}}$ and $\ket{\psi_{\mathsf{a}}}$ states, and introducing the aforementioned approximations, we get the following system of coupled differential equations (see Appendix \ref{App:Derivation:A} for a detailed derivation)
\begin{align}
    &i\hbar \dv{\ket{\Phi_{\mathsf{b}}(t)}}{t}
            = e \mel{\psi_{\mathsf{b}}}{\hat{\boldsymbol{R}}(t)}{\psi_{\mathsf{b}}}
                \cdot \hat{\boldsymbol{E}}(t)
                \ket{\Phi_{\mathsf{b}}(t)} \nonumber
            \\&\hspace{2.1cm} 
            + e \mel{\psi_{\mathsf{b}}}{\hat{\boldsymbol{R}}(t)}{\psi_{\mathsf{a}}}
            \cdot \hat{\boldsymbol{E}}(t)
        \ket{\Phi_{\mathsf{a}}(t)}, \label{Eq:system:A}
    \\
    &i\hbar \dv{\ket{\Phi_{\mathsf{a}}(t)}}{t}
        = e \mel{\psi_{\mathsf{a}}}{\hat{\boldsymbol{R}}(t)}{\psi_{\mathsf{b}}}
            \cdot \hat{\boldsymbol{E}}(t)
            \ket{\Phi_{\mathsf{b}}(t)} \nonumber
        \\&\hspace{2.1cm} 
            + e \mel{\psi_{\mathsf{a}}}{\hat{\boldsymbol{R}}(t)}{\psi_{\mathsf{a}}}
			\cdot \hat{\boldsymbol{E}}(t)
			\ket{\Phi_{\mathsf{a}}(t)}, \label{Eq:system:B}
\end{align}
where $\ket{\Phi_{\mathsf{i}}(t)} \equiv \langle \psi_{\mathsf{i}}\vert\Tilde{\Psi}(t)\rangle$, with $\lvert \Tilde{\Psi}(t)\rangle = \hat{U}_{\text{sc}}(t) \lvert \bar{\Psi}(t)\rangle$, is the quantum optical state when the electron is found in state $\ket{\psi_{\mathsf{i}}}$. Thus, in this expression, we have two different contributions: one given by $\boldsymbol{\mu}_{\mathsf{ii}}(t) \equiv e\mel{\psi_{\mathsf{i}}}{\hat{\boldsymbol{R}}(t)}{\psi_{\mathsf{i}}}$, i.e., the average time-dependent dipole moment with respect to state $\ket{\psi_{\mathsf{i}}}$; and a second one given by $\boldsymbol{\mu}_{\mathsf{ij}}(t) \equiv e\mel{\psi_{\mathsf{i}}}{\hat{\boldsymbol{R}}(t)}{\psi_{\mathsf{j}}}$, with $\mathsf{i}\neq \mathsf{j}$, which couples both differential equations.
	
A solution to the system of differential equations presented in Eqs.~\eqref{Eq:system:A} and \eqref{Eq:system:B}, with the initial conditions included, can be written as (see Appendix~\ref{App:Derivation:B})
\begin{equation}\label{Eq:Bonding:Recursive}
\begin{aligned}
    \ket{\Phi_{\mathsf{b}}(t)}
        &= \hat{\mathcal{D}}\big(\boldsymbol{\chi}_{\mathsf{b}}(t,t_0)\big)
                \ket{\bar{0}}
        \\& \quad
            -\dfrac{1}{\hbar^2} 
                \int^t_{t_0} \dd t_1  \int^{t_1}_{t_0}	\dd t_2
                    \hat{\mathcal{D}}\big(\boldsymbol{\chi}_{\mathsf{b}}(t,t_1)\big)
                    \hat{M}_{\mathsf{ba}}(t_1)
            \\&\hspace{2cm}
                \times
                    \hat{\mathcal{D}}\big(\boldsymbol{\chi}_{\mathsf{a}}(t,t_1)\big)
                        \hat{M}_{\mathsf{ab}}(t_2)\ket{\Phi_{\mathsf{b}}(t_2)},
    \end{aligned}
\end{equation}
for the bonding quantum optical component, i.e. when the electron is found in a bonding state, while for the antibonding term we get
\begin{equation}\label{Eq:Antibonding:Recursive}
    \begin{aligned}
        \ket{\Phi_{\mathsf{a}}(t)}
            &=
                - \dfrac{i}{\hbar}
                    \int^t_{t_0} \dd t_1
                        \hat{\mathcal{D}}(\boldsymbol{\chi}_{\mathsf{a}}(t,t_1))
                        \hat{M}_{\mathsf{a}\mathsf{b}}(t_1)
                    \\
                    &\hspace{2cm}\times
                        \hat{\mathcal{D}}(\boldsymbol{\chi}_{\mathsf{b}}(t_1,t_0))
                            \ket{\bar{0}}
                    \\&\quad
                        - \dfrac{1}{\hbar^2}
                            \int^t_{t_0} \dd t_1 \int^{t_1}_{t_0} \dd t_2
                                \hat{\mathcal{D}}(\boldsymbol{\chi}_{\mathsf{a}}(t,t_1))
                                \hat{M}_{\mathsf{a}\mathsf{b}}(t_1)
                                \\&\hspace{2cm}\times
                                \hat{\mathcal{D}}(\boldsymbol{\chi}_{\mathsf{b}}(t_1,t_2))
                                \hat{M}_{\mathsf{b}\mathsf{a}}(t_2)
                                \ket{\Phi_{\mathsf{a}}(t_2)}.
    \end{aligned}
\end{equation}

In Eqs~\eqref{Eq:Bonding:Recursive} and \eqref{Eq:Antibonding:Recursive}, we have that $\ket{\bar{0}} \equiv \bigotimes_{q,\mu}\ket{0_{q,\mu}}$, $\hat{M}_{\mathsf{ij}}(t) = e \boldsymbol{\mu}_{\mathsf{ij}}(t) \cdot \hat{\boldsymbol{E}}(t)$, $\hat{\mathcal{D}}(\boldsymbol{\chi}_{\mathsf{i}}) \equiv e^{i\varphi_{\mathsf{i}}(t)}\prod_{q} \hat{D}\big(\chi_{\mathsf{i}}^{(q)}\big)$ with $\hat{D}(\chi^{(q)}) = \exp[\chi^{(q)} \hat{a}_q^\dagger - (\chi^{(q)})^* \hat{a}_q]$ the displacement operator with respect to the $q$th field mode \cite{ScullyBook,Gerry__Book_2001} and $\varphi_{\mathsf{i}}(t)$ a phase factor (see Appendix \ref{App:Derivation:B} for details), and where
\begin{equation}
    \chi^{(q)}_{\mathsf{i}}(t,t_0)
        = -\dfrac{1}{\hbar}
            \int^t_{t_0} \dd \tau
                e^{i\omega_q \tau} \boldsymbol{\mu}_{{\mathsf{ii}}}(t) \cdot \vb{g}(\omega_q),
\end{equation} 
that is, the Fourier transform of the averaged time-dependent dipole moment with respect to the electronic state $\ket{\psi_{\mathsf{i}}}$.

Let us carefully analyze the processes described by these equations. In  Eq.~\eqref{Eq:Bonding:Recursive}, we have that the first term depicts a process where the only bound state the electron populates is the bonding state. As a consequence of the HHG dynamics, which are identical to those happening in atomic-HHG processes, each harmonic mode of the electromagnetic field gets shifted a quantity $\chi^{(q)}_{\mathsf{b}}(t,t_0)$ \cite{lewenstein_generation_2021,rivera-dean_strong_2022,stammer_quantum_2023}. The second term presents a process where at time $t_2$, the electron transitions from a bonding to an antibonding state, the interaction described by the $\hat{M}_{\mathsf{ab}}(t)$ operator. Between the time intervals $t_2$ and $t_1$, with $t_1 \geq t_2$, the field modes get displaced by $\chi^{(q)}_{\mathsf{a}}(t_1,t_2)$, as a consequence of the interaction of the electron with the field modes when it is located in the antibonding state. Finally, at time $t_1$ the electron returns to the bonding state, where it stays until the end of the pulse with the field modes getting displaced a quantity $\chi^{(q)}_{\mathsf{b}}(t,t_1)$. Similar dynamics are obtained for the antibonding state, with the main difference that the first term of Eq.~\eqref{Eq:Bonding:Recursive} is missing. This is a consequence of our initial conditions, Eq.~\eqref{Eq:Init:cond}, since we impose the electron to be at $t_0$ in a bonding state. Thus, the only way to find the electron in an antibonding state is by means of a transition from the bonding component. Apart from this difference, the analysis of Eq.~\eqref{Eq:Antibonding:Recursive} is analogous to the one we have just presented.

The solutions shown in Eqs.~\eqref{Eq:Bonding:Recursive} and \eqref{Eq:Antibonding:Recursive}, define a recursive relation for the bonding and antibonding quantum optical components. Each recursive iteration leads to an extra interaction between these two states. In the following, we truncate our equations up to first-order with respect to the interaction processes: we allow the electron to, at most, perform a single transition from the bonding or antibonding states. Note that this is valid under the regime $\abs{\boldsymbol{\mu}_{\mathsf{bb}}(t)} > \abs{\boldsymbol{\mu}_{\mathsf{ba}}(t)}$ ($\abs{\boldsymbol{\mu}_{\mathsf{aa}}(t)} > \abs{\boldsymbol{\mu}_{\mathsf{ba}}(t)}$), i.e., when the probability of performing a transition from a bonding to an antibonding (or vice versa) state is lower than the probability of staying in a bonding (antibonding) state. For the HHG processes, this is typically the situation, since the electron eventually ionizes from and recombines to the ground state. However, one could potentially alter this situation in molecular systems by using non-symmetric targets \cite{bian_multichannel_2010}, i.e.~diatomic molecules where the atoms in each center belong to different species, and/or by adding a {\it perturbative} ultraviolet field with a relative phase with respect to that of the intense infrared radiation \cite{bian_phase_2011}.

After this truncation, we rewrite Eqs.~\eqref{Eq:Bonding:Recursive} and \eqref{Eq:Antibonding:Recursive} as
\begin{align}
        &\ket{\Phi_{\mathsf{b}}(t)}
            \approx \hat{\mathcal{D}}
                            \big(
                                \boldsymbol{\chi}_{\mathsf{b}}(t,t_0)
                            \big)\bigotimes_{q,\mu} \ket{0_{q,\mu}}   \label{Eq:Bond:approx}
        \\
        &\ket{\Phi_{\mathsf{a}}(t)}
            \approx -\dfrac{i}{\hbar}
                    \int^t_{t_0} \dd t_1
                        \hat{\mathcal{D}}\big(\boldsymbol{\chi}_{\mathsf{a}}(t,t_1)\big)
                        \hat{\mathcal{D}}\big(\boldsymbol{\chi}_{\mathsf{b}}(t_1,t_0)\big)
                        \label{Eq:Antbond:approx}
                        \\&\hspace{2.5cm}\times
                            \boldsymbol{\mu}_{\mathsf{ab}}(t_1)
                            \cdot\big(
                                \hat{\boldsymbol{E}}(t_1)
                                + \boldsymbol{E}_{\text{cl}}^{(\mathsf{b})}(t_1)
                            \big)
                        \bigotimes_{q,\mu}\ket{0_{q,\mu}}
   \nonumber,
\end{align}
where in Eq.~\eqref{Eq:Antbond:approx} we have that $\boldsymbol{E}_{\text{cl}}^{(\mathsf{b})}(t) \equiv \mel{\boldsymbol{\chi}_{\mathsf{b}}(t)}{\hat{\boldsymbol{E}}(t)}{\boldsymbol{\chi}_{\mathsf{b}}(t)}$ (see Appendix \ref{App:Derivation:B} for a more detailed derivation).

With all this, we have that the final joint state for the electron and the electromagnetic field after HHG is approximately given by
\begin{equation}\label{Eq:HHG:Ant:Bond}
    \lvert \Tilde{\Psi}(t) \rangle
        \approx \dfrac{1}{\sqrt{\mathcal{N}}}
            \big[ 
                \ket{\psi_{\mathsf{b}}(t)}
                    \ket{\Phi_{\mathsf{b}}(t)}
                + 	\ket{\psi_{\mathsf{a}}(t)}
                        \ket{\Phi_{\mathsf{a}}(t)}
            \big],
\end{equation}
which, in general, has the form of an entangled state between the electronic and quantum optical degrees of freedom. Alternatively, one could also provide an interpretation of this state in terms of recombination events taking place in the right or left atomic centers. Note that, according to Ref.~\cite{lein_mechanisms_2005}, a transfer mechanism where the electron ionizes at one center and recombines in the other becomes efficient when the electron is initially in a delocalized state. This is the case of the ground (bonding) state of H$_2^+$ (see Fig.~\ref{Fig:Scheme}). Here, we introduce the set of \emph{localized} states $\{\Rbar,\Lbar\}$, given by $\Rbar = (1/\sqrt{2})[\ket{\psi_{\mathsf{b}}} + \ket{\psi_{\mathsf{a}}}]$ and $\Lbar = (1/\sqrt{2})[\ket{\psi_{\mathsf{b}}} - \ket{\psi_{\mathsf{a}}}]$, which unlike the set $\{\ket{\text{g}_R}, \ket{\text{g}_L}\}$, define an orthonormal set that is slightly more localized in the right and left centers compared to that of the atomic orbitals (solid blue and red curves in Fig.~\ref{Fig:Scheme}~(a)). In the limit when the distance between the two centers becomes infinitely large, both sets converge.

Under this \emph{localized right} and \emph{left} set, we can rewrite the state in Eq.~\eqref{Eq:HHG:Ant:Bond} as
\begin{equation}\label{Eq:HHG:R:L}
    \begin{aligned}
    \lvert \Tilde{\Psi}(t) \rangle
        &= \dfrac{1}{\sqrt{2\mathcal{N}}}
            \Big[
                \Rbar 
                \big(
                    \ket{\Phi_{\mathsf{b}}(t)}
                    +\ket{\Phi_{\mathsf{a}}(t)}
                \big)
                \\&\hspace{1.5cm}
                + \Lbar 
                \big(
                    \ket{\Phi_{\mathsf{b}}(t)}
                    -\ket{\Phi_{\mathsf{a}}(t)}
                \big)
            \Big]
        \\&=
            \dfrac{1}{\sqrt{2\mathcal{N}}}
            \big[
                \Rbar 
                    \ket{\Phi_{\bar{R}}(t)}
                + \Lbar 
                    \ket{\Phi_{\bar{L}}(t)}
            \big]
    \end{aligned}
\end{equation}
which presents the same amount of entanglement as Eq.~\eqref{Eq:HHG:Ant:Bond} since local unitary transformations leave the total amount of entanglement invariant \cite{nielsen_quantum_2010}. However, by performing measurements that are able to distinguish between the localized right and left components ($\hat{P}_{\bar{R}} = \dyad{\bar{R}},\hat{P}_{\bar{L}} = \dyad{\bar{L}}$), or to distinguish between different energetic states  ($\hat{P}_{\mathsf{b}} = \dyad{\psi_{\mathsf{b}}},\hat{P}_{\mathsf{a}} = \dyad{\psi_{\mathsf{a}}}$), the final quantum optical state gets modified, as it will be studied in the remnant of this work.

\section{RESULTS}\label{Sec:Results}
\begin{figure*}
    \centering
    \includegraphics[width=0.75\textwidth]{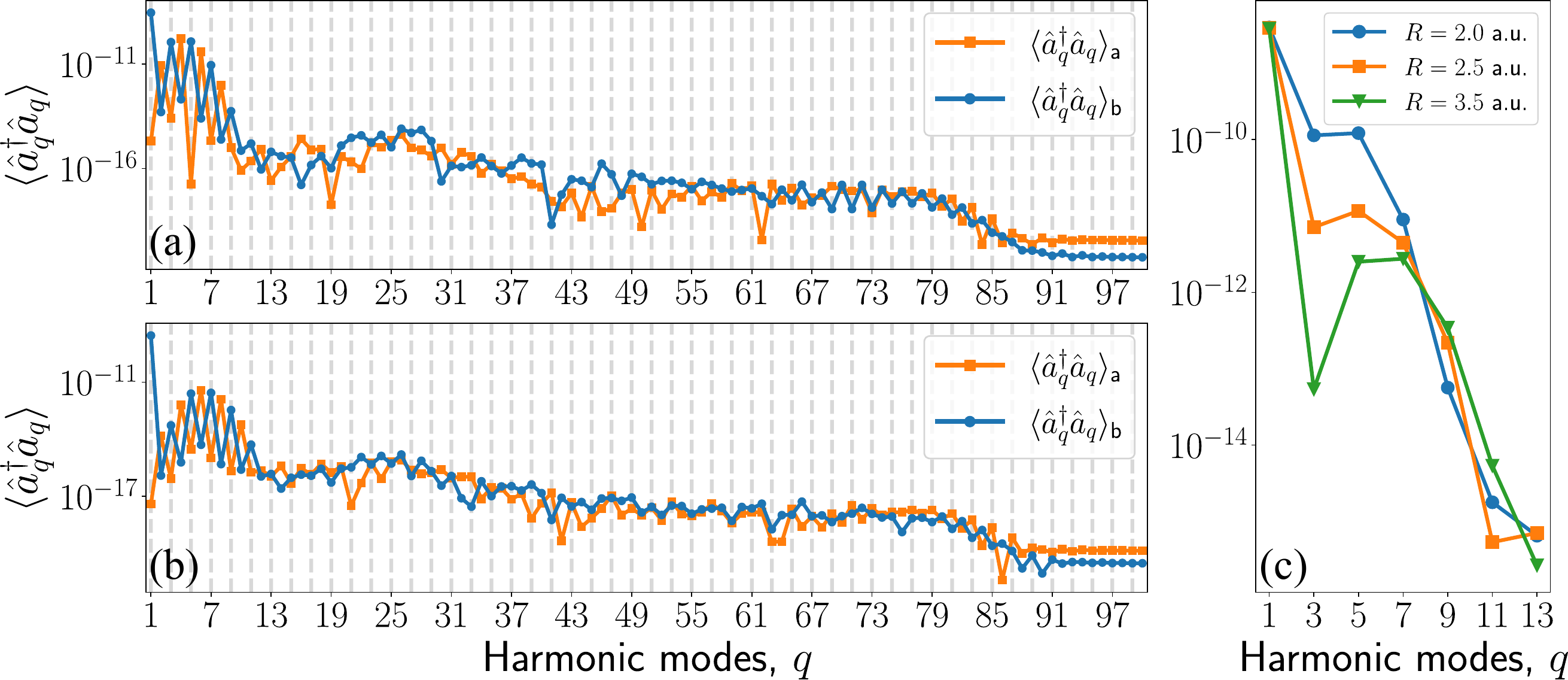}
    \caption{Mean photon number distribution for the different harmonic modes when considering single-molecule dynamics. In (a) and (b), the calculations have been done for interatomic distances $R=2.0$ a.u. and $R=3.5$ a.u., respectively, and the blue and orange curves correspond to the bonding and antibonding components. In (c), we show the mean photon number for odd harmonic orders, where each curve corresponds to a different interatomic distance. For these calculations, we have considered an H$_2^+$ molecule excited with a sinusoidal squared laser pulse of peak intensity $I=5 \times 10^{14}$ W/cm$^2$, central wavelength $\lambda_L = 800$ nm and $\Delta t \sim 21$ fs of duration (8 optical cycles).}
    \label{Fig:mpn:single}
\end{figure*}
In this section, we study different quantum optical and quantum information quantities of the states presented in Eqs.~\eqref{Eq:HHG:Ant:Bond} and \eqref{Eq:HHG:R:L}. For the numerical analysis, we consider that an $\text{H}_2^+$ molecular ion is driven by a $\sin^2$-envelope pulse linearly polarized along the molecular axis, with central wavelength $\lambda_L = 800$ nm, peak intensity $I=5\times 10^{14}$ W/cm$^2$ and a total duration of $\Delta t \approx 21$ fs (8 optical cycles).

\subsection{Mean photon number and the many-molecules regime}
Here, we look at the mean photon number of the different harmonic modes obtained from Eq.~\eqref{Eq:HHG:Ant:Bond} (or equivalently Eq.~\eqref{Eq:HHG:R:L}) which, to some extent, should resemble the harmonic spectra measured after HHG processes. This will allow us to benchmark the predictions of our theory from those obtained with semiclassical approaches \cite{suarez_high-order-harmonic_2017,suarez_rojas_strong-field_2018}. Ultimately, this comparison will be used to discuss a phenomenological many-molecule extension to the single-molecule calculations we have done thus far, although a more elaborated derivation of this is presented in Appendix \ref{App:Many:Mol}.

Assuming that we have no knowledge about what state has the electron recombined with, the quantum optical state reads
\begin{equation}\label{Eq:HHG:QO:Full}
    \begin{aligned}
    \hat{\rho}_f(t)
        &=\tr_{\text{elec}}
            \big(
                \lvert\Tilde{\Psi}(t)\rangle \!
                \langle \Tilde{\Psi}(t)\rvert
            \big)
        \\&
        = \dfrac{1}{\mathcal{N}}
            \big[
                \dyad{\Phi_{\mathsf{b}}(t)}
                + \dyad{\Phi_{\mathsf{a}}(t)}
            \big],
    \end{aligned}
\end{equation}
where we have performed the partial trace with respect to the electronic degrees of freedom. The mean photon number present in the $q$th harmonic mode is then given by
\begin{equation}\label{Eq:mpn}
    \expval{\hat{a}^\dagger_q \hat{a}_q}
        = \dfrac{1}{\mathcal{N}}
            \Big[
                    \expval{\hat{a}^\dagger_q \hat{a}_q}_{\mathsf{b}}
                    + \expval{\hat{a}^\dagger_q \hat{a}_q}_{\mathsf{a}}
            \Big],
\end{equation}
where we have defined $\expval{\hat{a}^\dagger_q \hat{a}_q}_{\mathsf{i}}\equiv \mel{\Phi_{\mathsf{i}}(t_{\text{end}})}{\hat{a}^\dagger_q\hat{a}_q}{\Phi_{\mathsf{i}}(t_{\text{end}})}$, with $t_{\text{end}}$ denoting the end of the pulse.

In Fig.~\ref{Fig:mpn:single} we show the results of this calculation when considering a single molecule interacting with the field. Specifically, in (a) and (b) we show separately the contributions of $\langle\hat{a}^\dagger_q \hat{a}_q\rangle_{\mathsf{b}}$ (blue curve with circular markers) and $\langle\hat{a}^\dagger_q \hat{a}_q\rangle_{\mathsf{a}}$ (orange curve with rectangular markers), for two different interatomic distances. In both cases, we recover some of the characteristic features of the $\text{H}_2^+$ HHG spectra \cite{lein_role_2002,suarez_rojas_strong-field_2018}: two plateau regions, a \emph{low-frequency} one happening between the 1st and the 13th harmonic, and a second for higher frequencies which lasts until the cutoff frequency, located around the 80th harmonic. 
While in the second plateau region, the presence of even and odd harmonics cannot be very well distinguished, a typical feature due to the interference between different electronic trajectories at recombination \cite{schafer_high_1997,salieres_temporal_1998}, for the first plateau region we can discern between two contributions: the $\langle\hat{a}^\dagger_q \hat{a}_q\rangle_{\mathsf{b}}$ term clearly contributes to odd harmonic orders, while $\langle\hat{a}^\dagger_q \hat{a}_q\rangle_{\mathsf{a}}$ to even harmonic orders. Before the HHG takes place, the electron is initially in a bonding state which, after recombination, ends up in an antibonding state that has opposite parity. This inversion of symmetry in the final electronic state, reflects in the mean photon number distribution with the presence of even harmonic orders \cite{bian_multichannel_2010,bian_phase_2011}.

\begin{figure}
    \centering
    \includegraphics[width=1\columnwidth]{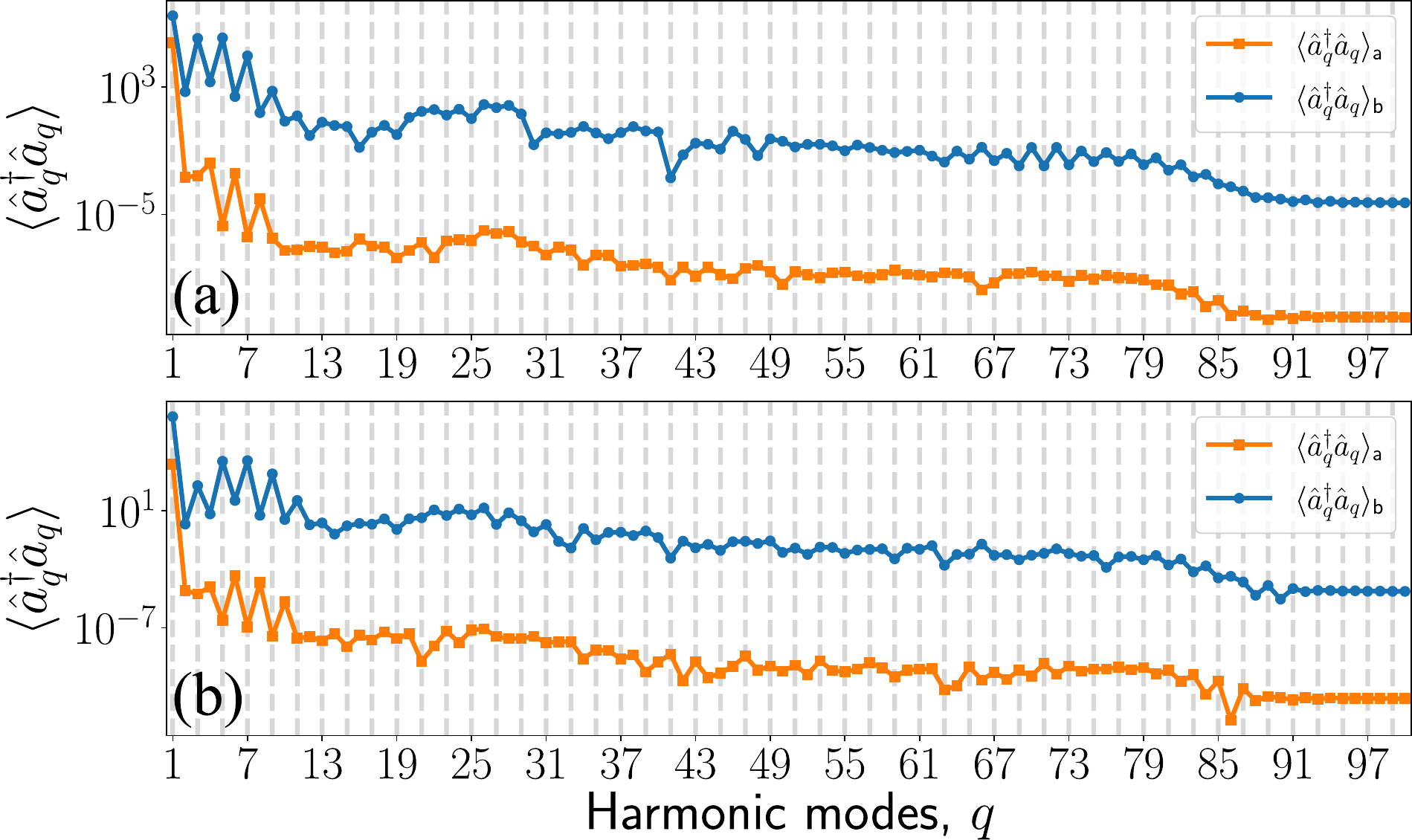}
    \caption{Mean photon number distribution for the different harmonic modes when considering $N_{\text{mol}}$ molecules \cite{note1}. Specifically, we set in both plots $N_{\text{mol}} = 10^8$, while  $R= 2.0$ a.u. in (a), and $R= 3.5$ a.u. in (b). We observe that recombination events ending up in a bonding state (blue curve with circular markers), provide a higher contribution to the mean photon number compared to those happening with an antibonding state (orange curve with squared markers). Specifically, the former scales with $N^2$ while the latter with $N$.}
    \label{Fig:mpn:many}
\end{figure}

One of the most surprising aspects about Figs.~\ref{Fig:mpn:single}~(a) and (b), is the relative contribution of $\expval{\hat{a}^\dagger_q \hat{a}_q}_{\mathsf{b}}$ and $\expval{\hat{a}^\dagger_q \hat{a}_q}_{\mathsf{a}}$, as they show the same order of magnitude. This is because we find that the probability of generating a photon, within the single-molecule scenario, is almost equal for the bonding-bonding and bonding-antibonding channels, ranging from $10^{-10}$ until $10^{-16}$ from the lowest to the highest harmonic orders in Fig.~\ref{Fig:mpn:single}~(a). On the other hand, when looking at the electronic population for both energetic states, we find that the probability of finding an electron in a bonding state is dominant, as in most cases the electron barely interacts with the field. We now provide an extension of our equations to the case where we have a system composed of $N_{\text{mol}}$ uncorrelated molecules interacting with the field~\cite{note1}. In order to do this, we take into account that in the many-molecule scenario, there are two different contributions to the measured HHG signal \cite{eberly_spectrum_1992}: a coherent contribution that scales as $N_{\text{mol}}^2$ coming from events where the electron recombines with the state from which it has ionized, and an incoherent contribution that scales as $N_{\text{mol}}$ from electrons that recombine with other bound states. Thus, one could phenomenologically take this into account by redefining the time-dependent dipole moments (for a more detailed derivation see Appendix~\ref{App:Many:Mol}). Specifically, we define the $N_{\text{mol}}$-time-dependent dipole moments as $\boldsymbol{\mu}^{(N_{\text{mol}})}_{\mathsf{bb}}(t) \equiv N_{\text{mol}} \boldsymbol{\mu}_{\mathsf{bb}}(t)$ and $\boldsymbol{\mu}^{(N_{\text{mol}})}_{\mathsf{ij}}(t) \equiv \sqrt{N_{\text{mol}}} \boldsymbol{\mu}_{\mathsf{ij}}(t)$ when $\mathsf{i}\neq\mathsf{j}$. By doing this, we get the mean photon number distribution shown in Fig.~\ref{Fig:mpn:many}, where we observe that the final mean photon number shows clear odd-harmonic orders along the first plateau region. In this case, one can check that $\langle \hat{a}_q^\dagger \hat{a}_q\rangle_{\mathsf{b}}$ scales with $N_{\text{mol}}^2$ while $\langle \hat{a}_q^\dagger \hat{a}_q\rangle_{\mathsf{a}}$ as $N_{\text{mol}}$.

To conclude with this section, let us discuss how increasing the interatomic distance affects the final mean photon number distribution of the harmonic modes. In Fig.~\ref{Fig:mpn:single}~(c), we show the total mean photon number (Eq.~\eqref{Eq:mpn}) for the odd-harmonic orders in the first harmonic plateau when considering three different interatomic distances. We observe that for larger distances, the peak of the harmonic spectrum for $q>1$ becomes smaller. This is better understood by considering a description of the HHG process in terms of recombinations with right and left centers. Under this picture, the larger the distance between the two centers, the lower the probability of ionization-recombination events taking place between different centers. Consequently, a lower efficiency of the HHG conversion is expected. However, it is important to note that the characteristics of the HHG spectrum can be modified when considering different molecular-field orientations as the interatomic distance varies \cite{han_internuclear-distance_2013}.

\subsection{Wigner function distribution}
One of the most complete ways of characterizing a quantum optical state is the Wigner function, as it encodes in phase-space all the information about it \cite{wigner_quantum_1932,Schleich_Book_2001}. Specifically, it has been widely used in the field of quantum optics as a witness of non-classical features, which are typically related to the presence of negative regions in the observed distribution and/or non-Gaussian behaviors \cite{hudson_when_1974,smithey_measurement_1993}. Following Ref.~\cite{royer_wigner_1977}, the Wigner function for the $q$th harmonic mode is proportional to the mean value of the operator $\hat{\mathcal{W}}(\beta) = \hat{D}_q(\beta)\hat{\Pi}_q\hat{D}^\dagger_q(\beta)$, with $\hat{\Pi}_q$ the parity operator acting on mode $q$.

\begin{figure}
    \centering
    \includegraphics[width=1\columnwidth]{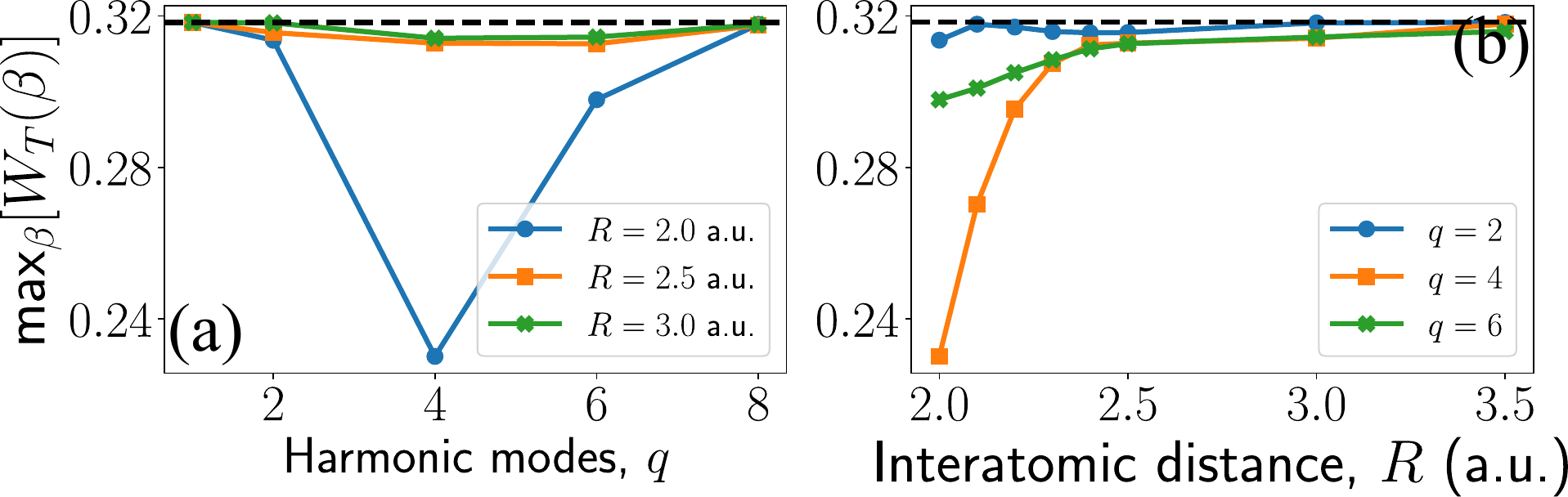}
    \caption{Maximum value of Wigner function of the state presented in Eq.~\eqref{Eq:HHG:QO:Full} in phase space as a function of: (a) the harmonic modes $q$ when considering different interatomic distances, and (b) the interatomic distance when considering different harmonic orders. In these plots, we have set $N_{\text{mol}} = 10^9$. Although not shown here, it was observed that the obtained Wigner functions presented a Gaussian-like behaviour.}
    \label{Fig:Wigner:tot}
\end{figure}

\begin{figure*}
    \centering
    \includegraphics[width=0.75\textwidth]{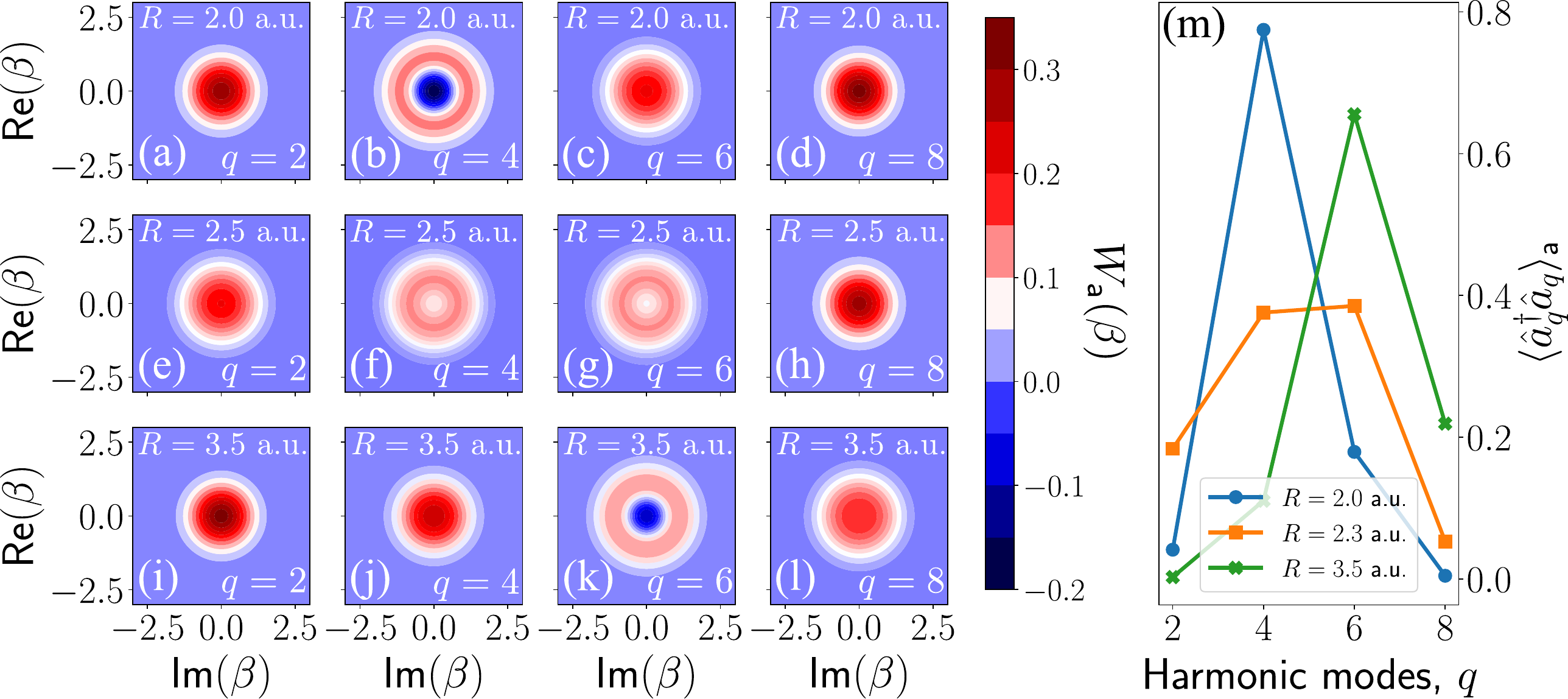}
    \caption{Wigner function of the quantum optical state when conditioning the electron to be found in an antibonding, i.e., $W_{\mathsf{a}}(\beta)$ in Eq.~\eqref{Eq:Wigner:tot}. Here, $\beta = \Tilde{\beta} - \alpha - \chi_{\mathsf{a}}(t)$. In (a)-(l), each of the rows correspond to a different interatomic distance ($R=2.0$ a.u., $2.5$ a.u. and $3.5$ a.u. from up to bottom), while each of the columns correspond to different harmonic modes ($q=2,4,6$ and 8 from left to right). In (m), we show the dependence of the mean photon number, computed with the antibonding component (once normalized to unity), with the harmonic modes and for different interatomic distances $R$.}
    \label{Fig:Wigner:ant}
\end{figure*}

In this section, we study the Wigner distribution of the quantum optical state after the HHG interaction under different circumstances. First, we consider the case of Eq.~\eqref{Eq:HHG:QO:Full}, where we have no knowledge about the final state of the electron. In this case, the Wigner function can be written as
\begin{equation}\label{Eq:Wigner:tot}
    \begin{aligned}
    W^{(q)}_T
        &= \dfrac{1}{\pi}\tr(\rho_f(t) \hat{\mathcal{W}}_q(\beta))
        \\&=\dfrac{1}{\mathcal{N}}
            \Big[
                W_{\mathsf{b}}^{(q)}(\beta)
                + \mathcal{N}_\mathsf{a}
                    W_{\mathsf{a}}^{(q)}(\beta)
            \Big],
    \end{aligned}
\end{equation}
where we have defined $W^{(q)}_{\mathsf{i}}(\beta) = \mel{\Phi_{\mathsf{i}}(t)}{\hat{\mathcal{W}}_q(\beta)}{\Phi_{\mathsf{i}}(t)}$ and $\mathcal{N}_{\mathsf{a}}$ is the normalization constant of $\ket{\Phi_{\mathsf{a}}(t)}$. Under the regimes we have studied, i.e. with the excitation conditions specified at the beginning of this section and for $N_{\text{mol}}\leq 10^9$, Eq.~\eqref{Eq:Wigner:tot} presented in all cases a Gaussian-like behavior. Yet, some differences are observed with the standard Wigner function observed for coherent states. By definition, the Wigner function of a coherent state $\ket{\alpha}$ is a Gaussian with maximum value equal to $\pi^{-1}$. However, because of the influence of the antibonding component in Eq.~\eqref{Eq:Wigner:tot} which depends on the number of molecules $N_{\text{mol}}$, we observe that the maximum value of these Wigner functions gets reduced, as is shown in Fig.~\ref{Fig:Wigner:tot}. Specifically, in Fig.~\ref{Fig:Wigner:tot}~(a) we show the dependence of this maximum value with respect to even harmonic orders. We specifically choose these values because, as shown in Figs.~\ref{Fig:mpn:single} and \ref{Fig:mpn:many}, these are the harmonic orders to which the antibonding quantum optical component contributes the most. Therefore, the higher the contribution of the corresponding quantum optical component to the state, the more affected we expect the maximum of the obtained Wigner distribution to be. On the other hand, the interatomic distance also plays a fundamental role, as observed in Fig.~\ref{Fig:Wigner:tot}~(b). Here, we see that the maximum of the Wigner function tends to $\pi^{-1}$ as $R$ increases. Specifically, the bigger $R$ is the less likely is to have ionization and recombination events between different centers. This translates into a lower occupation of the antibonding state and, hence, into a smaller variation of the Wigner function maxima.

In Refs.~\cite{lewenstein_generation_2021,rivera-dean_strong_2022,stammer_high_2022,stammer_theory_2022,stammer_quantum_2023}, the presence of non-classical states of light were observed in atomic, and recently in solid-state \cite{gonoskov_nonclassical_2022,rivera-dean_entanglement_2023}, systems upon the performance of quantum operations that restrict to instances where high-order harmonic radiation is generated. From an experimental perspective, this requires the performance of a(n) (anti)correlated measurement between the generated harmonics and part of the fundamental mode \cite{tsatrafyllis_high-order_2017}. Here, instead of performing this kind of conditioning operations, we constrain our state to those instances where the electron has ended up in an antibonding state after HHG. Mathematically, this corresponds to the case where we apply the projector $\hat{P}_{\mathsf{a}} = \dyad{\psi_{\mathsf{a}}}$ onto Eq.~\eqref{Eq:HHG:Ant:Bond}, such that the resulting quantum optical state is $\ket{\Phi_{\mathsf{a}}(t)}$. In a similar basis to what has been found in the aforementioned references, one could expect to observe non-classical features in this case as well.

In Fig.~\ref{Fig:Wigner:ant}~(a)-(l), we show the obtained Wigner function for different harmonic modes and for different interatomic distances. More specifically, we have $R=2.0$ a.u. for the first row, $R=2.5$ a.u. for the second and the $R=3.5$ a.u for the third. In most cases, we observe that the Wigner functions of the different harmonic modes present a Gaussian-like behavior, with a more or less flat maximum depending on the harmonic mode. However, in some cases, the Wigner function shows a distinctive ring-like shaped distribution (see Fig.~\ref{Fig:Wigner:ant}~(b),~(f),~(g) and (k)). When comparing these results with the corresponding mean-photon number distribution, shown in Fig.~\ref{Fig:Wigner:ant}~(m), we see that for the cases where the ring-like-shaped distribution is observed, the mean-photon number reaches its maximum value. This is related to the fact that, for these harmonic modes, the highest contribution to the state $\rho^{(q)}_{\mathsf{a}} = \tr_{q'\neq q} (\dyad{\Phi_{\mathsf{a}}(t)})$ comes from (displaced) single-photon states (see Appendix~\ref{App:Many:Mol}). The bigger is the contribution of the single-photon state, the more profound is the obtained central minima. Further note, that in this analysis we have omitted odd harmonic orders, as the recombination events we are looking at, for small harmonic orders, generate photons at even harmonic orders (see Figs.~\ref{Fig:mpn:single} and \ref{Fig:mpn:many}).

Finally, to conclude this section, we note that if the same analysis is done when considering the recombination process that either ends in the right or left atomic centers ($\lvert\bar{R}\rangle$ or $\lvert\bar{L}\rangle$), we get similar results to those found in Fig.~\ref{Fig:Wigner:tot}. Here, the resulting quantum optical states are given as the superposition of the bonding and antibonding quantum optical components, the former being dominant over the latter, leading to Gaussian-like Wigner functions.

\subsection{Electron-light entanglement}
As we have seen in the previous section, the final state of the electron plays a crucial role in determining the features observed in the quantum optical state. Thus, given the structure of the state shown in Eq.~\eqref{Eq:HHG:Ant:Bond} (equivalently Eq.~\eqref{Eq:HHG:R:L}), one could expect the electronic and field degrees of freedom to be entangled. The use of this kind of hybrid entangled states \cite{van_loock_optical_2011}, has proven to be extremely useful for different quantum information science tasks, such as quantum teleportation \cite{he_teleportation_2022}, quantum communication \cite{zhang_device-independent_2022}, quantum steering \cite{cavailles_demonstration_2018} and fault-tolerant quantum computing \cite{omkar_resource-efficient_2020}. Therefore, given that HHG processes could provide access to this kind of states \cite{rivera-dean_light-matter_2022,rivera-dean_entanglement_2023}, here we study the light-matter entanglement between the electronic and electromagnetic field degrees of freedom. Since we are dealing with pure states, we can characterize the entanglement features of the obtained state by means of the entropy of entanglement, i.e., $S(\hat{\sigma}) := - \tr(\hat{\sigma}\log_2\hat{\sigma})$ \cite{plenio_introduction_2007,amico_entanglement_2008,nielsen_quantum_2010}. In this definition, $\hat{\sigma}$ corresponds to the reduced density matrix with respect to one of the subsystems. For the sake of simplicity, in our calculations, we perform the partial trace with respect to the electromagnetic field degrees of freedom, such that we use $\hat{\sigma} \equiv \hat{\rho}_{\text{elec}}(t) = \tr_f(\lvert \Tilde{\Psi}(t) \rangle\!\langle\Tilde{\Psi}(t)\rvert)$ (see Appendix~\ref{App:light:matter}).
\begin{figure}
    \centering
    \includegraphics[width=1\columnwidth]{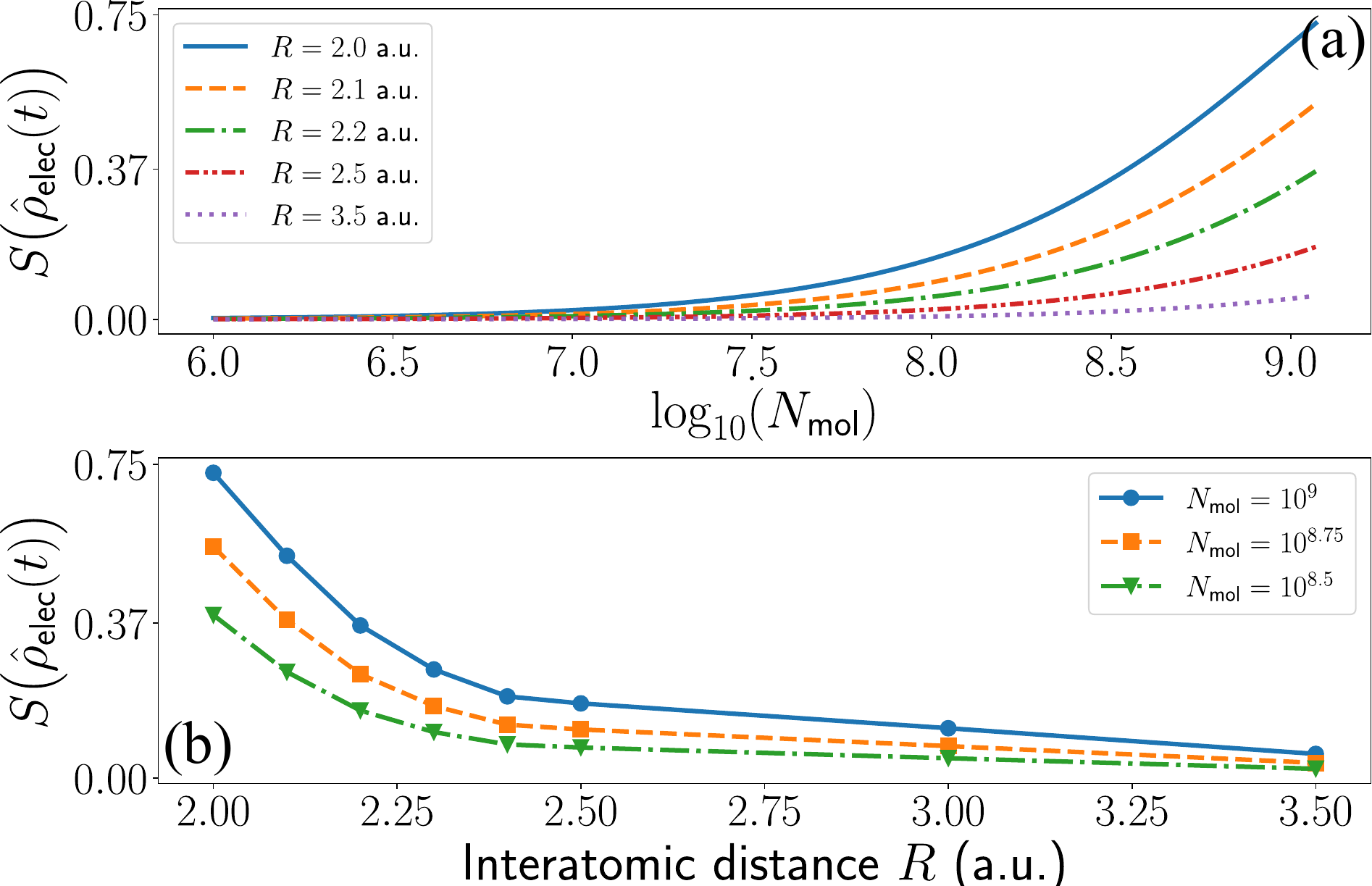}
    \caption{Light-matter entanglement between the electronic degrees of freedom and all the harmonic modes. In (a), we show the dependence of the entropy of entanglement as a function of the number of molecules, when considering different interatomic distances $R$. In (b), we instead fix the number of molecules, and show the dependence of $S(\hat{\rho}_{\text{elec}}(t))$ as a function of the interatomic distance.}
    \label{Fig:light:elec:ent}
\end{figure}

In Fig.~\ref{Fig:light:elec:ent}, we present the obtained results. In (a), we observe that $S(\hat{\rho}_{\text{elec}}(t))$ increases exponentially with the number of interacting molecules, with the rate being determined by the distance $R$ between the atomic centers. Specifically, the maximum amount of entanglement is found to be $S(\hat{\rho}_{\text{elec}}(t))\approx 0.75$ for $R=2.0$ a.u. and $N_{\text{mol}} = 1.2 \times 10^9$. In (b), we instead observe for fixed values of $N_{\text{mol}}$, that $S(\hat{\rho}_{\text{elec}}(t))$ decreases for increasing interatomic distances. This is a consequence of the fact that, for small values of $R$, it is more likely to find processes where an electron ionizes from one center and recombines at the other, which ultimately enhances the probability of ending up in an antibonding state. We note that, in this treatment, the entanglement occurs between the field and the molecular ions and not between the molecules themselves. Furthermore, the values of $N_{\text{mol}}$ we use here match the estimates found in experiments with atomic gases \cite{lewenstein_generation_2021}, once including the proper correction factors \cite{note1}, which suggests that these effects could be potentially observed experimentally. 

\subsection{Entanglement between the harmonic modes}
\begin{figure}
    \centering
    \includegraphics[width=1\columnwidth]{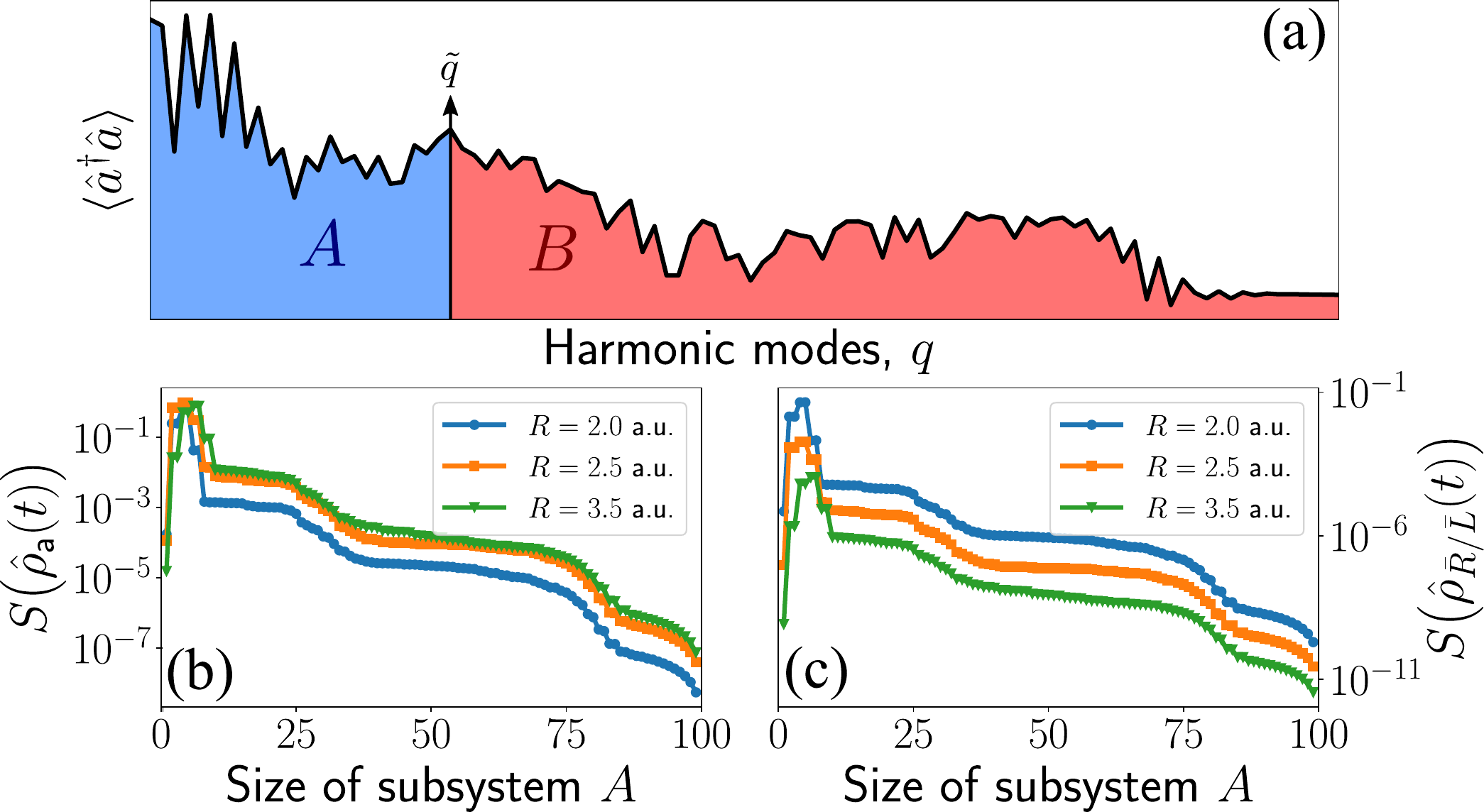}
    \caption{Entanglement between the field modes when the electron is conditioned to be on a specific quantum state. In (a), we show a schematic of the approach. We split the harmonic modes into two sets, $A := \{q : q \leq \Tilde{q}\}$ and $B := \{q:q> \Tilde{q}\}$, and study the entanglement between them as a function of the subsystem size, which we refer in the text as $\norm{A}$. In (b), we show how the entropy of entanglement $S(\hat{\rho}_{\mathsf{a}}(t))$ changes when the electron is conditioned to be found in an antibonding state, when considering different interatomic distances. In (c), we show the entropy of entanglement $S(\hat{\rho}_{\bar{R}/\bar{L}}(t))$ in the limit of having a large number of molecules, computed for the case where we condition the electron to be found in the localized right ($\lvert \bar{R}\rangle$) or left ($\lvert \bar{L}\rangle$) states.}
    \label{Fig:light:light:ent:cond}
\end{figure}

High-harmonic generation processes allow for the generation of light with frequencies spanning from the infrared to the extreme ultraviolet regime (see Figs.~\ref{Fig:mpn:single} and \ref{Fig:mpn:many}). This unique feature, together with the so-called \emph{conditioning on HHG} approaches \cite{lewenstein_generation_2021,rivera-dean_strong_2022,stammer_theory_2022}, allows for the generation of massively frequency-entangled light states \cite{stammer_high_2022,stammer_theory_2022,stammer_quantum_2023}, which could be of potential interest towards optical-based quantum information science applications \cite{peacock_time_2016,sciara_scalable_2021,lewenstein_attosecond_2022,bhattacharya_strong_2023}. In this section, we study the entanglement between two distinct sets of frequency modes under different scenarios. First, we consider the case where the electron is found in a given state, and then divide the frequency modes into two sets, $A := \{q: q \leq \Tilde{q}\}$ and $B := \{q: q > \Tilde{q}\}$ (see Fig.~\ref{Fig:light:light:ent:cond}~(a)), and study the amount of entanglement between the two sets. Then, we consider the case where we have no knowledge about what state the electron ends up in, and characterize the entanglement between one of the frequency modes and the rest. Note that, in general, one could consider more general entanglement characterizations involving more than two parties, which is a topic of active research \cite{walter_multipartite_2016,bengtsson_brief_2016}. Here, we restrict to bipartite scenarios for which general entanglement measures and witnesses are well-known \cite{amico_entanglement_2008}. 

For the first case, we condition the electron state to be found either in an antibonding ($\ket{\psi_{\mathsf{a}}}$), a left ($\lvert \bar{L}\rangle$) or a right ($\lvert \bar{R}\rangle$) state. Thus, the quantum optical state once considering the separation between partitions $A$ and $B$, can be generally written as
\begin{equation}\label{Eq:ent:ent:charact}
    \begin{aligned}
    \ket{\Phi_\mathsf{i}}
        &= a_{\mathsf{i}} 
            \hat{\mathcal{D}}(\Tilde{\boldsymbol{\chi}})
                \ket{\Tilde{0}_A}\ket{\Tilde{0}_B}
         + b_{\mathsf{i}}
            \ket{\Tilde{1}_A}\ket{\Tilde{0}_B}
        \\&\quad
        + c_{\mathsf{i}}
            \ket{\Tilde{0}_A}\ket{\Tilde{1}_B}
        + d_{\mathsf{i}}
            \ket{\Tilde{0}_A}\ket{\Tilde{0}_B},
    \end{aligned}
\end{equation}
with $\mathsf{i} = \{\mathsf{a},\bar{R},\bar{L}\}$ such that the coefficients $\{a_{\mathsf{i}},b_{\mathsf{i}},c_{\mathsf{i}},d_{\mathsf{i}}\}$ depend on the state with respect to which we have projected the electronic part (see Appendix \ref{App:light:light:ent}). Specifically, when considering the projection with respect to an antibonding state, we get that $a_{\mathsf{a}} = 0$, and we can easily characterize the amount of entanglement by looking at the entropy of entanglement. The results for this case are shown in Fig.~\ref{Fig:light:light:ent:cond}~(b) as a function of the size of subsystem $A$, denoted hereupon as $\norm{A}= \dim(A)$, for different interatomic distances. We observe that there is an optimal value of $\norm{A}$ for which the entanglement achieves a maximum value. Specifically, we find that, for  $R=2.0, 2.5$ and $3.5$ a.u., we respectively get $\max_{\norm{A}}[S(\hat{\rho}_{\mathsf{a}}(t))] \approx 0.69, 0.99$ and $0.78$. Thus, we see that by properly defining sets $A$ and $B$, one can generate highly frequency-entangled bipartite states. We note that the values of $\norm{A}$ for which $S(\hat{\rho}_{\mathsf{a}}(t))$ becomes maximum, corresponds to those definitions of $A$ (equivalently $B$) for which $\Tilde{q}$ is a harmonic mode with a maximum value of the mean-photon number (see Figs.~\ref{Fig:mpn:single}~(a) and (b)). When increasing $\norm{A}$ beyond this value, the entropy of entanglement gets reduced following the typical plateau-like structure of usual HHG spectra: the probability of generating a photon in a mode belonging to subsystem $B$ becomes less likely when higher harmonic orders are included in this set, which reduces the quantum correlations between both subsystems. We also observe that for larger interatomic distances $R$, the amount of entanglement for $\norm{A} \gtrsim 30$ increases. This is because the harmonic yield decreases for larger interatomic distances, which makes the difference in population between the low and high harmonic regimes smaller. Nevertheless, we emphasize that the probability of having recombination events with an antibonding state becomes smaller as the interatomic distance becomes larger. 

On the other hand, when conditioning the electron to be found either in the right or left centers, the amount of entanglement shows a similar behaviour to the one we discussed (see Fig.~\ref{Fig:light:light:ent:cond}~(c)), although leading to smaller values of entanglement. Specifically, the maximum values that we find for for $R=2.0$ a.u., $R=2.5$ a.u. and $R=3.5$ a.u., are $\max_{\norm{A}}[S(\hat{\rho}_{\bar{R}/\bar{L}}(t))] \approx 4.4\times 10^{-2}, 1.8\times 10^{-3}$ and $1.11 \times 10^{-4}$, respectively. Thus, conditioning the electron to be either found in the localized right or left components, hugely influences the final amount of entanglement, compared to the case where we condition the electron to be found in an antibonding state. We also note that, for increasing interatomic distances, the amount of entanglement reduces for all possible values of $\norm{A}$, which is in stark contrast with what is observed in Fig.~\ref{Fig:light:light:ent:cond}~(b). This is an expected feature, since the amount of entanglement obtained when projecting onto the localized right and/or left basis, is strongly influenced by a drift of population from the bonding and antibonding states. Thus, the more likely is to find this kind of transitions, which in particular occurs for small interatomic distances, the more entangled will be subsystems $A$ and $B$.

\begin{figure}
    \centering
    \includegraphics[width=1\columnwidth]{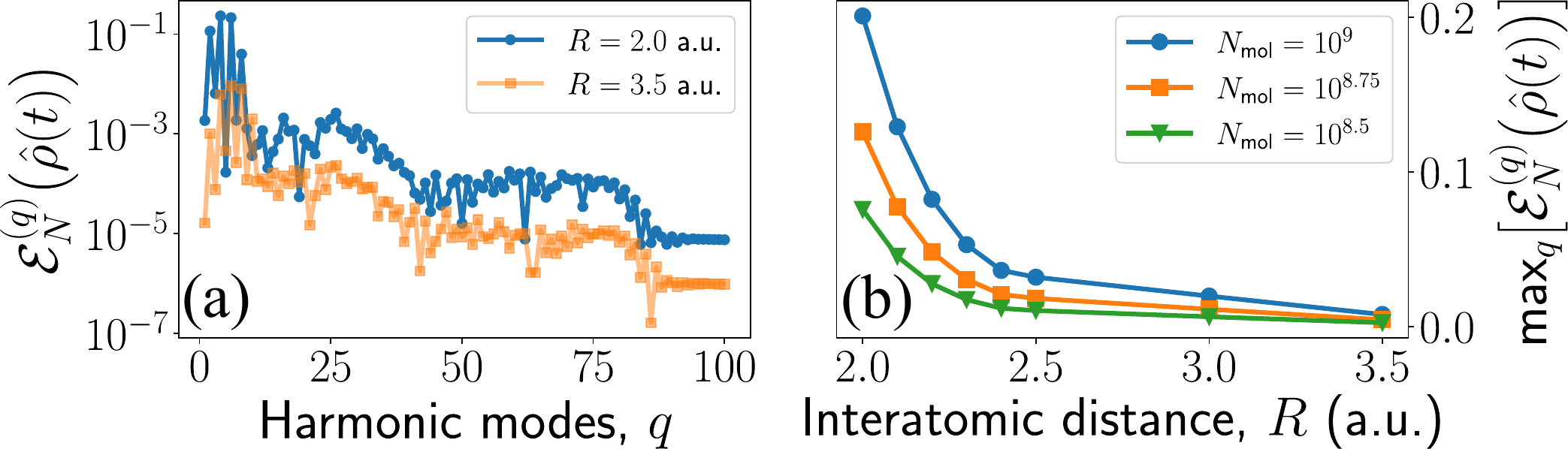}
    \caption{Entanglement between the field modes assuming that we have no knowledge about the electronic state. Here we study the amount of entanglement between a single mode $\Tilde{q}$ and the rest. Thus, instead of working with the sets $A$ and $B$ (see Fig.~\ref{Fig:light:light:ent:cond}), we instead have  $\bar{A} := \{\Tilde{q}\}$ and $\bar{B} := \{q: \forall q \neq \Tilde{q}\}$. In (a), we show the amount of entanglement between the $q$th mode and the rest, when considering different interatomic distances and $N_{\text{mol}} = 10^8$. As a measure of entanglement, we use a lower bound on the logarithmic negativity $\mathcal{E}^{(q)}_N(\hat{\rho}(t))$ (see the text for more details). In (b), we compute $\max_q[ \mathcal{E}^{(q)}_N(\hat{\rho}(t))]$ for different number of molecules, and show its dependence with $R$.}
    \label{Fig:light:light:ent:LN}
\end{figure}

Alternatively, we study the amount of entanglement between the $q$th harmonic and the rest, in the case where we have no knowledge about the final state of the electron. Thus, in this case, we work with the quantum optical state given by Eq.~\eqref{Eq:HHG:QO:Full}, i.e.,  a mixed state and for which entropy of entanglement is not a valid entanglement measure \cite{amico_entanglement_2008}. Instead, we can use entanglement witnesses such as the logarithmic negativity, which witnesses the presence of non-PPT (Positive Partial Transpose) entangled states, and that is defined as \cite{vidal_computable_2002,plenio_logarithmic_2005}
\begin{equation}\label{Eq:Log:Neg}
    E_N(\hat{\rho})
        :=\log_2(2 \mathsf{N}+1),
\end{equation}
where $\mathsf{N}$ is the negativity, that is, the sum of all negative eigenvalues of the partial transpose of $\hat{\rho}$ with respect to one of the subsystems. Since we are dealing with displaced Fock states, the calculation of  Eq.~\eqref{Eq:Log:Neg} becomes computationally demanding. In order to overcome this, we propose the following lower bound to the logarithmic negativity (see Appendix \ref{App:light:light:ent})
\begin{equation}
    E_N(\hat{\rho}_f(t)) 
        \geq 
            \mathcal{E}_N(\hat{\rho}_f(t))
                = \log_2
                    \Big(
                        2\abs{\min_i \lambda^{T_{\Tilde{B}}}_{\mathsf{a},i}}+1
                    \Big),
\end{equation}
where $\{\lambda^{T_{\Tilde{B}}}_{\mathsf{a},i}\}$ are the eigenvalues of $\hat{\rho}^{T_{\Tilde{B}}}_{\mathsf{a}}(t) \propto(\dyad{\Phi_{\mathsf{a}}(t)})^{T_{\bar{B}}}$, where $T_{\bar{B}}$ denotes the partial transpose with respect to subsystem $\Tilde{B}$, defined as $\bar{B} := \{q: \forall q\neq \Tilde{q}\}$ (and with $\bar{A} := \{\Tilde{q}\}$).

In Fig.~\ref{Fig:light:light:ent:LN}~(a) we show the results of this computation as a function of the harmonic modes, and for two interatomic distances. In the low harmonic regime ($q \lesssim 15$), this entanglement measure shows peaks for even harmonic orders, while troughs for the odd harmonic orders. We note that the presence of entanglement in this state is influenced by the presence of recombination events that end up in an antibonding state. For these, we have observed that, within the range $q \lesssim 15$, even orders are the ones that get populated the most, and in some cases, they lead to non-classical signatures in their Wigner function distribution (see Fig.~\ref{Fig:Wigner:ant}). Therefore, it is reasonable that for these modes, $\mathcal{E}_N(\hat{\rho}_f(t))$ becomes maximum ($\sim 0.2$ for $R=2.0$ a.u and $\sim9.29\times 10^{-3}$ for $R=3.5$ a.u.). When increasing $q$  beyond the 15th harmonic, $\mathcal{E}_N(\hat{\rho}_f(t))$ shows similar features to that of the harmonic spectrum: a second plateau region which extends until cutoff, after which the entanglement measure shows an abrupt decrease. On the other hand, larger interatomic distances $R$ lead to smaller values of $\mathcal{E}_N(\hat{\rho}_f(t))$. This is better observed in Fig.~\ref{Fig:light:light:ent:LN}~(b), where we show  $\max_q[\mathcal{E}_N(\hat{\rho}_f(t))]$ as a function of $R$.

\section{CONCLUSIONS AND OUTLOOK}\label{Sec:Conclusions}
In this study, we have undertaken a theoretical investigation of the high-order harmonic generation (HHG) process in H$_2^+$ molecular ions within a quantum optical framework. Our research has focused on characterizing various quantum optical and quantum information meaures, revealing the versatility of HHG in two-center molecules for quantum technology applications. We have demonstrated the emergence of entanglement between the electron and light states following their interaction, leading to the generation of hybrid-entangled states. These states hold significant relevance in the advancement of quantum technology applications, as emphasized throughout this work.

Furthermore, we have identified that, by selectively examining events where the electron ends up in specific quantum states, it becomes possible to obtain non-classical states of light in targeted frequency modes. Additionally, we have achieved the generation of highly entangled states between distinct sets of harmonic modes.

It is worth noting that our study was conducted under specific conditions, focusing on symmetric two-center molecules aligned along the polarization axis of the incident laser field. We anticipate that the observed features will exhibit strong dependence on (i) the polarization and orientation of the laser field with respect to the molecular axis, as well as the molecular alignment, as they crucially affect the harmonic emission (see for instance \cite{suarez_high-order-harmonic_2017,suarez_rojas_strong-field_2018} and references therein), and (ii) the population of different bound states and electron localization, which are heavily influenced by the molecular structure encompassing the atomic species within each center and the interatomic separation distance \cite{bian_multichannel_2010,bian_phase_2011,lein_mechanisms_2005}. To further exploit the potential of the interface between quantum optics and HHG in diatomic molecules toward quantum technology applications, future investigations should consider these factors.

As a possible outlook, it would be interesting to extend the present theory to multi-center and large molecules as, for instance, those employed in the semiclassical limit in the series of papers \cite{suarez_high-order-harmonic_2017,suarez_above-threshold_2016,suarez_above-threshold_2018} and PhD  thesis \cite{suarez_rojas_strong-field_2018} by Noslen Suárez.

\section*{ACKNOWLEDGMENTS}
We acknowledge insightful discussions with A.~F.~Ordóñez and H.~B.~Crispin.

ICFO group acknowledges support from: ERC AdG NOQIA; Ministerio de Ciencia y Innovation Agencia Estatal de Investigaciones (PGC2018--097027--B--I00 / 10.13039 / 501100011033, CEX2019--000910--S / 10.13039 / 501100011033, Plan National FIDEUA PID 2019--106901GB--I00, FPI, QUANTERA MAQS PCI 2019--111828--2, QUANTERA DYNAMITE PCI 2022--132919,  Proyectos de I+D+I “Retos Colaboración” QUSPIN RTC 2019--007196--7); MICIIN with funding from European Union Next Generation EU (PRTR--C17.I1) and by Generalitat de Catalunya;  Fundació Cellex; Fundació Mir-Puig; Generalitat de Catalunya (European Social Fund FEDER and CERCA program, AGAUR Grant No. 2021 SGR 01452, QuantumCAT \ U16--011424, co-funded by ERDF Operational Program of Catalonia 2014-2020); Barcelona Supercomputing Center MareNostrum (FI--2022--1--0042); EU Horizon 2020 FET--OPEN OPTOlogic (Grant No 899794); EU Horizon Europe Program (Grant Agreement 101080086 — NeQST), National Science Centre, Poland (Symfonia Grant No. 2016/20/W/ST4/00314); ICFO Internal “QuantumGaudi” project; European Union’s Horizon 2020 research and innovation program under the Marie-Skłodowska-Curie grant agreement No 101029393 (STREDCH) and No 847648  (“La Caixa” Junior Leaders fellowships ID100010434 : LCF / BQ / PI19 / 11690013, LCF / BQ / PI20 / 11760031, LCF / BQ / PR20 / 11770012, LCF / BQ / PR21 / 11840013). Views and opinions expressed in this work are, however, those of the author(s) only and do not necessarily reflect those of the European Union, European Climate, Infrastructure and Environment Executive Agency (CINEA), nor any other granting authority.  Neither the European Union nor any granting authority can be held responsible for them. 

P. Tzallas group at FORTH acknowledges LASERLABEUROPE V (H2020-EU.1.4.1.2 grant no.871124), FORTH Synergy Grant AgiIDA (grand no. 00133), the H2020 framework program for research and innovation under the NEP-Europe-Pilot project (no. 101007417). ELI-ALPS is supported by the European Union and co-financed by the European Regional Development Fund (GINOP Grant No. 2.3.6-15-2015-00001).

J.R-D. acknowledges funding from the Secretaria d'Universitats i Recerca del Departament d'Empresa i Coneixement de la Generalitat de Catalunya, the European Social Fund (L'FSE inverteix en el teu futur)--FEDER, the Government of Spain (Severo Ochoa CEX2019-000910-S and TRANQI), Fundació Cellex, Fundació Mir-Puig, Generalitat de Catalunya (CERCA program) and the ERC AdG CERQUTE. 

P.S. acknowledges funding from the European Union’s Horizon 2020 research and innovation programme under the Marie Skłodowska-Curie grant agreement No 847517. 

A.S.M. acknowledges funding support from the European Union’s Horizon 2020 research and innovation programme under the Marie Sk{\l}odowska-Curie grant agreement, SSFI No.\ 887153.

E.P. acknowledges the Royal Society for University Research Fellowship funding under URF$\setminus$R1$\setminus$211390.

M.F.C. acknowledges financial support from the Guangdong Province Science and Technology Major Project (Future functional materials under extreme conditions - 2021B0301030005) and the Guangdong Natural Science Foundation (General Program project No. 2023A1515010871).

\bibliography{References.bib}{}

\begin{thebibliography}{99}
\bibitem [{not()}]{note1}%
  \BibitemOpen
  \href@noop {} {\bibinfo {title} {{In our numerical calculations, the
  parameter $N_{\text{mol}}$ does not correspond exactly with the number of
  molecules, but it is proportional to it. This is because when computing
  numerically the time-dependent dipole moments, the results we get are in
  arbitrary units. Thus, due to the use of these arbitrary units, there is a
  missing factor between the actual number of molecules and the
  $N_{\text{mol}}$ quantity we introduce. From our calculations, we estimate
  these missing factors to be in the order of $10^{-2}\sim
  10^{-3}$.}}}\BibitemShut {Stop}%
\end{thebibliography}
\clearpage
\onecolumngrid
\appendix

\begin{center}
\large{\textbf{\textsc{Appendix}}}
\end{center}
\section{Analysis of the Time-Dependent Schrödinger equation}\label{App:First}
In this appendix, we aim to give a more detailed derivation on how we solved the Schrödinger equation, and the approximations we have considered, in order to reach Eq.~\eqref{Eq:HHG:Ant:Bond} of the main text. 

\subsection{Presenting the equations}\label{App:Derivation:A}
We start our discussion with Eq.~\eqref{Eq:Sch:before} after moving to the interaction picture with respect to the semiclassical Hamiltonian $H_{\text{sc}}(t) = \hat{H}_{\text{mol}} + e \hat{\boldsymbol{R}}\cdot \boldsymbol{E}_{\text{cl}}(t)$. Under this picture, the position operator $\hat{\boldsymbol{R}}$ acquires a time-dependence, i.e., $\hat{\boldsymbol{R}}(t) \equiv \hat{U}^\dagger_{\text{sc}}(t) \hat{\boldsymbol{R}}\hat{U}_{\text{sc}}(t)$ with $\hat{U}_{\text{sc}}(t) = \hat{\mathcal{T}}e^{-\tfrac{i}{\hbar}\int^t_{t_0} \dd \tau \hat{H}_{\text{sc}}(\tau)}$, where $\hat{\mathcal{T}}$ is the time-ordering operator. Thus, the resulting Schrödinger equation reads
\begin{equation}\label{Eq:App:Schrod:after}
    i \hbar \pdv{\vert\Tilde{\Psi}(t)\rangle}{t}
        = e \hat{\boldsymbol{R}}(t)\cdot \hat{\boldsymbol{E}}(t)
        \vert\Tilde{\Psi}(t)\rangle,
\end{equation}
where $\vert\Tilde{\Psi}(t)\rangle = \hat{U}_{\text{sc}}(t) \ket{\bar{\Psi}(t)}$. We now introduce the identity in the electronic subspace as
\begin{equation}
    \mathbbm{1}
        = \dyad{\psi_0} + \dyad{\psi_1}
        + \sum_{n=2} \dyad{\psi_n}
        + \int \dd \psi_c \dyad{\psi_c},
\end{equation}
where the first two terms are the projectors with respect to the ground and first excited state of the molecule, which we assume are not degenerate; the third term contains the projector that includes all other bound states, and the last one all those related to the continuum states. Inserting this expression in Eq.~\eqref{Eq:App:Schrod:after}, we get
\begin{equation}\label{Eq:App:Sch:Identity}
    i\hbar \pdv{\vert\Tilde{\Psi}(t)\rangle}{t}
        = e \hat{\boldsymbol{R}}(t)\cdot \hat{\boldsymbol{E}}(t)
        \bigg[
            \langle \psi_0\vert\Tilde{\Psi}(t)\rangle
                \ket{\psi_0}
            + \langle \psi_1\vert\Tilde{\Psi}(t)\rangle
                \ket{\psi_1}
            + \sum_{n=2} \langle \psi_n\vert\Tilde{\Psi}(t)\rangle
                \ket{\psi_n}
            + \int \dd \psi_c
                \langle \psi_c\vert\Tilde{\Psi}(t)\rangle
                    \ket{\psi_c}
        \bigg].
\end{equation}

Similarly to what has been done for atomic-HHG processes \cite{lewenstein_generation_2021,rivera-dean_strong_2022,stammer_quantum_2023}, we neglect the electronic continuum population at all times \cite{stammer_quantum_2023,stammer_theory_2022}, as this contribution is typically considered to be small in comparison to that of the molecular lowest energy states \cite{lewenstein_theory_1994,suarez_high-order-harmonic_2017,suarez_rojas_strong-field_2018}. Also, we consider a slightly different version of the Strong-Field Approximation (SFA) \cite{lewenstein_theory_1994}. In the SFA, one of the assumptions is that the only bound state contributing to the dynamics is the ground state. On top of this one, here we also consider the contribution of the first excited state of the molecule. And this is because, for the case we are interested in, i.e., H$_2^+$ molecular ions, these two states are crucial for spanning a set of localized states, which determines whether the electron is closer to the atom on the right or the atom on the left of the considered diatomic molecule (see Fig.~\ref{Fig:Scheme} for a pictorial representation).

Thus, we approximate Eq.~\eqref{Eq:App:Sch:Identity} as
\begin{equation}
    i\hbar \pdv{\vert\Tilde{\Psi}(t)\rangle}{t}
        \approx e \hat{\boldsymbol{R}}(t)\cdot \hat{\boldsymbol{E}}(t)
        \Big[
            \langle \psi_0\vert\Tilde{\Psi}(t)\rangle
                \ket{\psi_0}
            + \langle \psi_1\vert\Tilde{\Psi}(t)\rangle
                \ket{\psi_1}
        \Big],
\end{equation}
and projecting the whole equation both with respect to the ground and the first excited state, we get the following system of coupled differential equations
\begin{align}
    & i\hbar \dv{\ket{\Phi_{0}(t)}}{t}
        = e \mel{\psi_0}{\hat{\boldsymbol{R}}(t)}{\psi_0}\cdot \hat{\boldsymbol{E}}(t) \ket{\Phi_0(t)}
        + e \mel{\psi_0}{\hat{\boldsymbol{R}}(t)}{\psi_1}\cdot \hat{\boldsymbol{E}}(t) \ket{\Phi_1(t)}\label{Eq:App:Sch:ground:first:1},
    \\
    & i\hbar \dv{\ket{\Phi_{1}(t)}}{t}
        = e \mel{\psi_1}{\hat{\boldsymbol{R}}(t)}{\psi_0}\cdot \hat{\boldsymbol{E}}(t) \ket{\Phi_0(t)}
        + e \mel{\psi_1}{\hat{\boldsymbol{R}}(t)}{\psi_1}\cdot \hat{\boldsymbol{E}}(t) \ket{\Phi_1(t)}\label{Eq:App:Sch:ground:first:2},
\end{align}
where we have defined $\ket{\Phi_i(t)} \equiv \langle\psi_i\vert \Tilde{\Psi}(t)\rangle$.

There are several methods for computing the ground and first excited states of molecules. Currently, this is an active field of research, especially for large molecules which show a large degree of correlation. Here, since we want to have an analysis where we can distinguish the localized contributions of the molecule, we opt for the Linear Combination of Atomic Orbital (LCAO) method \cite{atkins_book,finkelstein_uber_1928}. According to this, the ground state molecular orbitals are expanded by linear combinations of atomic orbitals. This method is particularly useful when considering simple molecules for which a small number of atomic orbitals provide a good description of the ground and first excited states, as it happens with H$_2^+$. In this case, the ground and first excited are referred to as bonding and antibonding states and, within the LCAO, are respectively given by $\ket{\psi_{\mathsf{b}}} \propto \ket{\text{g}_L} + \ket{\text{g}_R}$ and $\ket{\psi_{\mathsf{a}}} \propto \ket{\text{g}_L} - \ket{\text{g}_R}$, with $\lvert\text{g}_{L}\rangle$ and $\lvert\text{g}_{R}\rangle$ the ground state orbitals of the atoms in the left ($L$) and right ($R$), respectively. Note that, as the number of atoms participating in the molecules grows larger, more atomic orbitals would be needed and the LCAO ceases to provide a straightforward description.

In terms of the bonding and antibonding states, Eqs.~\eqref{Eq:App:Sch:ground:first:1} and \eqref{Eq:App:Sch:ground:first:2} read
\begin{align}
    & i\hbar \dv{\ket{\Phi_{\mathsf{b}}(t)}}{t}
        = e \mel{\psi_{\mathsf{b}}}{\hat{\boldsymbol{R}}(t)}{\psi_{\mathsf{b}}}\cdot \hat{\boldsymbol{E}}(t) \ket{\Phi_{\mathsf{b}}(t)}
        + e \mel{\psi_{\mathsf{b}}}{\hat{\boldsymbol{R}}(t)}{\psi_{\mathsf{a}}}\cdot \hat{\boldsymbol{E}}(t) \ket{\Phi_{\mathsf{a}}(t)},
        \label{Eq:App:Sch:bond:antbond:1}
    \\
    & i\hbar \dv{\ket{\Phi_{\mathsf{a}}(t)}}{t}
        = e \mel{\psi_{\mathsf{a}}}{\hat{\boldsymbol{R}}(t)}{\psi_{\mathsf{b}}}\cdot \hat{\boldsymbol{E}}(t) \ket{\Phi_{\mathsf{b}}(t)}
        + e \mel{\psi_{\mathsf{a}}}{\hat{\boldsymbol{R}}(t)}{\psi_{\mathsf{a}}}\cdot \hat{\boldsymbol{E}}(t) \ket{\Phi_{\mathsf{a}}(t)}, \label{Eq:App:Sch:bond:antbond:2}
\end{align}
which are the equations we are going to focus on hereupon.

\subsection{Solving the equations in the single-molecule scenario}\label{App:Derivation:B}
The system of equations defined by Eqs.~\eqref{Eq:App:Sch:bond:antbond:1} and \eqref{Eq:App:Sch:bond:antbond:2}, define a system of coupled differential equations. If we take a closer look at these equations, one can see that both of them are first-order inhomogeneous differential equations with a well-defined homogeneous and inhomogeneous parts. Thus, their respective solution can be written as a sum of a solution to the homogeneous part, plus a particular solution to the inhomogeneous equation. Then, a solution to these equations can be written as
\begin{align}
    &\ket{\Phi_{\mathsf{b}}(t)}
        = \hat{\mathcal{D}}\big(\boldsymbol{\chi}_{\mathsf{b}}(t,t_0)\big)
            \ket{\Phi_{\mathsf{b}}(t,t_0)}
            - \dfrac{i}{\hbar}\int^t_{t_0} \dd t_1 
                \hat{\mathcal{D}}\big(\boldsymbol{\chi}_{\mathsf{b}}(t,t_1)\big)
                \hat{M}_{\mathsf{ba}}(t_1) \ket{\Phi_{\mathsf{a}}(t_1)}
                \label{Eq:App:sol:bond}
    \\
    &\ket{\Phi_{\mathsf{a}}(t)}
        = \hat{\mathcal{D}}\big(\boldsymbol{\chi}_{\mathsf{a}}(t,t_0)\big)
            \ket{\Phi_{\mathsf{a}}(t,t_0)}
            - \dfrac{i}{\hbar}\int^t_{t_0} \dd t_1 
                \hat{\mathcal{D}}\big(\boldsymbol{\chi}_{\mathsf{a}}(t,t_1)\big)
                \hat{M}_{\mathsf{ab}}(t_1) \ket{\Phi_{\mathsf{b}}(t_1)},
                \label{Eq:App:sol:antbond}
\end{align}
where in this expression we have defined $\hat{M}_{\mathsf{ij}} = e\mel{\psi_{\mathsf{i}}}{\hat{\boldsymbol{R}}(t)}{\psi_{\mathsf{j}}}\cdot \hat{\boldsymbol{E}}(t)$, $\hat{\mathcal{D}}(\boldsymbol{\chi}_{\mathsf{i}}) \equiv e^{i\varphi_{\mathsf{i}}(t)}\prod_q \hat{D}(\chi_{\mathsf{i}}^{(q)})$, with $\hat{D}(\chi^{(q)}) = \exp[\chi^{(q)}\hat{a}^\dagger_q - (\chi^{(q)})^*\hat{a}_q]$ the displacement operator acting on the $q$th field mode \cite{ScullyBook,Gerry__Book_2001}, where the phase factor $\varphi_{\mathsf{i}}(t)$ and the displacement $\chi^{(q)}_{\mathsf{i}}(t,t_0)$ are given by \cite{rivera-dean_strong_2022,stammer_quantum_2023}  
\begin{equation}
    \begin{aligned}
    &\varphi_{\mathsf{i}}(t) 
    = \dfrac{e^2}{\hbar^2}
        \sum_q \int^t_{t_0} \dd t_1 \int^{t_1}_{t_0} \dd t_2
            \big[
                \vb{g}(\omega_q)\cdot \boldsymbol{\mu}_{\mathsf{ii}}(t_1)
            \big]
            \big[
                \vb{g}(\omega_q)\cdot \boldsymbol{\mu}_{\mathsf{ii}}(t_2)
            \big]\sin(\omega_q(t_1-t_2)),
    \\
    & \chi_{\mathsf{i}}^{(q)}(t,t_0)
        = -\dfrac{1}{\hbar}
            \int^t_{t_0} \dd \tau
                e^{i\omega_q\tau}
                \boldsymbol{\mu}_{\mathsf{ii}}(\tau)
                \cdot \vb{g}(\omega_q),
    \end{aligned}
\end{equation}
noting that in these expressions we have further defined $\boldsymbol{\mu}_{\mathsf{ij}}(t) \equiv \mel{\psi_{\mathsf{i}}}{\hat{\boldsymbol{R}}(t)}{\psi_{\mathsf{j}}}$.

As we can see see, the solution for each equation depends on the other, as expected from the coupled structure of the considered system of equations. In fact, by introducing \eqref{Eq:App:sol:bond} inside \eqref{Eq:App:sol:antbond}, and vice versa, one gets a set of recursive relations
\begin{align}
        &\ket{\Phi_{\mathsf{b}}(t)}
            = \hat{\mathcal{D}}\big(\boldsymbol{\chi}_{\mathsf{b}}(t,t_0)\big)
                \ket{\Phi_{\mathsf{b}}(t,t_0)}
                - \dfrac{i}{\hbar}\int^t_{t_0} \dd t_1 
                    \hat{\mathcal{D}}\big(\boldsymbol{\chi}_{\mathsf{b}}(t,t_1)\big)
                    \hat{M}_{\mathsf{ba}}(t_1) 
                    \hat{\mathcal{D}}\big(\boldsymbol{\chi}_{\mathsf{a}}(t_1,t_0)\big)
                    \ket{\Phi_{\mathsf{a}}(t_0)}\nonumber
            \\&\hspace{5cm}
                + \bigg(\dfrac{i}{\hbar}\bigg)^2
                    \int^t_{t_0} \dd t_1
                        \int^{t_1}_{t_0} \dd t_2
                        \hat{\mathcal{D}}\big(\boldsymbol{\chi}_{\mathsf{b}}(t,t_1)\big)
                        \hat{M}_{\mathsf{ba}}(t_1)
                        \hat{\mathcal{D}}\big(\boldsymbol{\chi}_{\mathsf{a}}(t_1,t_2)\big)
                        \hat{M}_{\mathsf{ab}}(t_2)
                            \ket{\Phi_{\mathsf{b}}(t_2)},
            \label{Eq:App:sol:bond:recurs}
        \\
        &\ket{\Phi_{\mathsf{a}}(t)}
            = \hat{\mathcal{D}}\big(\boldsymbol{\chi}_{\mathsf{a}}(t,t_0)\big)
                \ket{\Phi_{\mathsf{a}}(t,t_0)}
                - \dfrac{i}{\hbar}\int^t_{t_0} \dd t_1 
                    \hat{\mathcal{D}}\big(\boldsymbol{\chi}_{\mathsf{a}}(t,t_1)\big)
                    \hat{M}_{\mathsf{ab}}(t_1) 
                    \hat{\mathcal{D}}\big(\boldsymbol{\chi}_{\mathsf{b}}(t_1,t_0)\big)
                    \ket{\Phi_{\mathsf{b}}(t_0)}\nonumber
            \\&\hspace{5cm}
                + \bigg(\dfrac{i}{\hbar}\bigg)^2
                    \int^t_{t_0} \dd t_1
                        \int^{t_1}_{t_0} \dd t_2
                        \hat{\mathcal{D}}\big(\boldsymbol{\chi}_{\mathsf{a}}(t,t_1)\big)
                        \hat{M}_{\mathsf{ab}}(t_1)
                        \hat{\mathcal{D}}\big(\boldsymbol{\chi}_{\mathsf{b}}(t_1,t_2)\big)
                        \hat{M}_{\mathsf{ba}}(t_2)
                            \ket{\Phi_{\mathsf{a}}(t_2)},
            \label{Eq:App:sol:antbond:recurs}
\end{align}
and as mentioned in the main text, each iteration of these recursive relations explicitly introduces higher-order transitions between the bonding and antibonding components from the initial state at $t_0$. Note that, in our case, we assume that initially the electron is located in the ground (bonding) state and the field in a vacuum state (within the displaced quantum-optical frame). Thus, we have that $\ket{\Phi_{\mathsf{b}}(t_0)} = \bigotimes_q\ket{0_q} \equiv \ket{\bar{0}}$ and $\ket{\Phi_{\mathsf{a}}(t_0)} = 0$, which introduced in \eqref{Eq:App:sol:bond:recurs} and \eqref{Eq:App:sol:antbond:recurs} leads to the recursive expressions shown in the main text
\begin{align}
    &\ket{\Phi_{\mathsf{b}}(t)}
            = \hat{\mathcal{D}}\big(\boldsymbol{\chi}_{\mathsf{b}}(t,t_0)\big)
                \ket{\bar{0}}
                + \bigg(\dfrac{i}{\hbar}\bigg)^2
                    \int^t_{t_0} \dd t_1
                        \int^{t_1}_{t_0} \dd t_2
                        \hat{\mathcal{D}}\big(\boldsymbol{\chi}_{\mathsf{b}}(t,t_1)\big)
                        \hat{M}_{\mathsf{ba}}(t_1)
                        \hat{\mathcal{D}}\big(\boldsymbol{\chi}_{\mathsf{a}}(t_1,t_2)\big)
                        \hat{M}_{\mathsf{ab}}(t_2)
                            \ket{\Phi_{\mathsf{b}}(t_2)},
            \label{Eq:App:sol:bond:recurs:t0}
        \\
        &\ket{\Phi_{\mathsf{a}}(t)}
            =  - \dfrac{i}{\hbar}\int^t_{t_0} \dd t_1 
                    \hat{\mathcal{D}}\big(\boldsymbol{\chi}_{\mathsf{a}}(t,t_1)\big)
                    \hat{M}_{\mathsf{ab}}(t_1)
                    \hat{\mathcal{D}}\big(\boldsymbol{\chi}_{\mathsf{b}}(t_1,t_0)\big)
                    \ket{\bar{0}}\nonumber
            \\&\hspace{1.5cm}
                + \bigg(\dfrac{i}{\hbar}\bigg)^2
                    \int^t_{t_0} \dd t_1
                        \int^{t_1}_{t_0} \dd t_2
                        \hat{\mathcal{D}}\big(\boldsymbol{\chi}_{\mathsf{a}}(t,t_1)\big)
                        \hat{M}_{\mathsf{ab}}(t_1)
                        \hat{\mathcal{D}}\big(\boldsymbol{\chi}_{\mathsf{b}}(t_1,t_2)\big)
                        \hat{M}_{\mathsf{ba}}(t_2)
                            \ket{\Phi_{\mathsf{a}}(t_2)}.
            \label{Eq:App:sol:antbond:recurs:t0}
\end{align}

\begin{figure}
    \centering
    \includegraphics[width=0.7\textwidth]{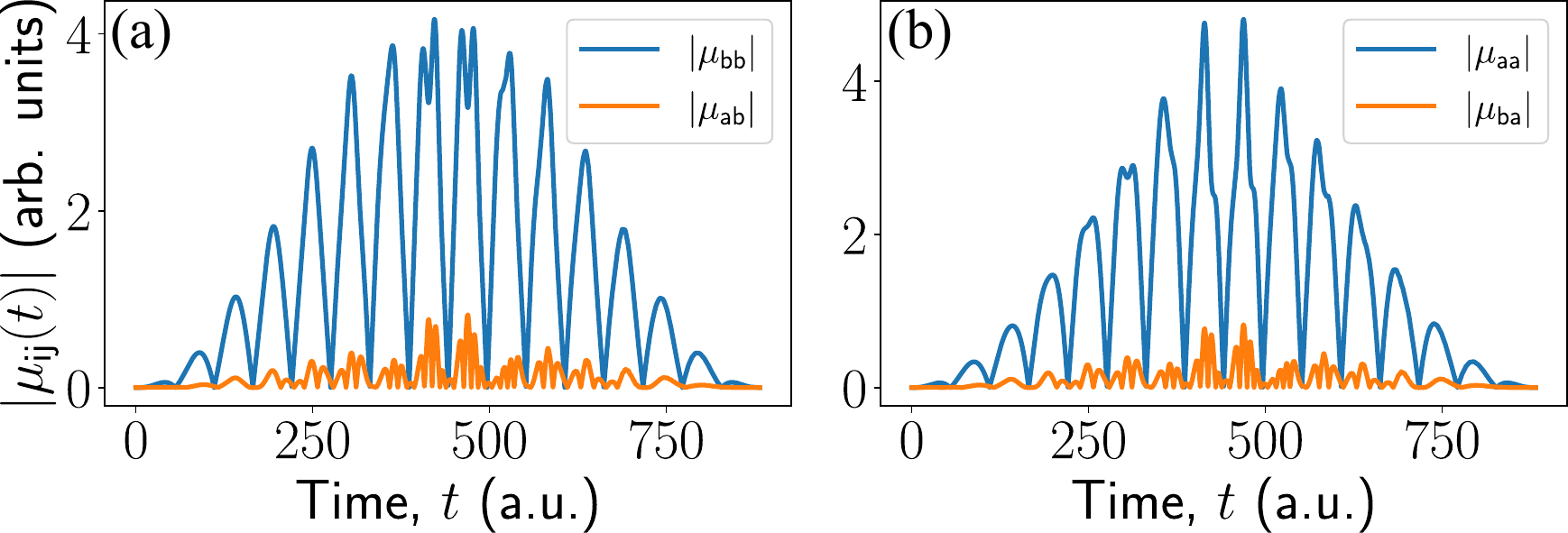}
    \caption{Different matrix elements of the time-dependent dipole moment for the case of $R=2.5$ a.u., under the excitation conditions we have worked with throughout the main text (see caption of Fig.~\ref{Fig:mpn:single} for instance).}
    \label{Fig:App:dipoles}
\end{figure}

As mentioned earlier, each iteration in the recursive relation introduces a new interaction between the bonding and antibonding states. In the following, we restrict our equations to the first-order interaction terms, i.e., we only allow the electron to perform a single transition from a bonding to an antibonding state. This approximation is valid in the regime $\abs{\mu_{\textsf{bb}}(t)} > \abs{\mu_{\textsf{ba}}(t)}$ and $\abs{\mu_{\textsf{aa}}(t)} > \abs{\mu_{\textsf{ba}}(t)}$ (note that by definition $\abs{\mu_{\textsf{ba}}(t)} = \abs{\mu_{\textsf{ab}}(t)}$), i.e., when at all integration times the probability (although in this case we express it in terms of the square root of the probability) of performing a transition from the bonding to an antibonding state (or vice versa) is lower than the probability of staying in a bonding (antibonding state). In Fig.~\ref{Fig:App:dipoles} we show that the conditions above are satisfied at all times. Here, we have considered the case of $R= 2.5$ a.u., although this election is arbitrary since a similar behaviour is observed for the range of interatomic distances considered in this article. The only difference between them is that for increasing values of $R$, we get that the relative difference between $\abs{\mu_{\mathsf{bb}}(t)}$ and $\abs{\mu_{\mathsf{ab}}(t)}$ (the same applies for $\abs{\mu_{\mathsf{aa}}(t)}$), becomes greater.

Therefore, up to first order, we approximate Eqs.~\eqref{Eq:App:sol:bond:recurs:t0} and \eqref{Eq:App:sol:antbond:recurs:t0} by
\begin{align}
    &\ket{\Phi_{\mathsf{b}}(t)}
        \approx \hat{\mathcal{D}}\big(\boldsymbol{\chi}_{\mathsf{b}}(t,t_0)\big)
            \ket{\bar{0}},
        \label{Eq:App:sol:bond:approx:t0}
    \\
    &\ket{\Phi_{\mathsf{a}}(t)}
        \approx 
        - \dfrac{i}{\hbar}\int^t_{t_0} \dd t_1 
                \hat{\mathcal{D}}\big(\boldsymbol{\chi}_{\mathsf{a}}(t,t_1)\big)
                \hat{M}_{\mathsf{ab}}(t_1)
                \hat{\mathcal{D}}\big(\boldsymbol{\chi}_{\mathsf{b}}(t_1,t_0)\big)
                \ket{\bar{0}}.
        \label{Eq:App:sol:antbond:approx:t0}
\end{align}

\subsection{Equations for the many-molecule scenario}\label{App:Many:Mol}
In the main text, we phenomenologically treated the many-molecule scenario by multiplying the time-dependent dipole moment matrix elements of the form $\boldsymbol{\mu}_{\mathsf{ii}(t)}$ by the number $N_{\text{mol}}$ of molecules, while the $\boldsymbol{\mu}_{\mathsf{ij}(t)}$ with $\mathsf{i}\neq \mathsf{j}$ by $\sqrt{N_{\text{mol}}}$. This picture was motivated by the one we have provided in Ref.~\cite{eberly_spectrum_1992}. In this section, we want to include a more elaborate basis for this phenomenological treatment.

Let us consider the case where we have $N_{\text{mol}}$ independent molecules excited by the driving field. In this case, the Hamiltonian of this system can be written as
\begin{equation}
    \hat{H} = \sum_{i=1}^{N_{\text{mol}}}
        \big[
            \hat{H}_{i,\text{mol}}
            + \hat{H}_{i,\text{int}}
        \big]
    + \hat{H}_{\text{field}},
\end{equation}
under the Born-Oppenheimer and dipole approximations, where the $i$ index runs through all the possible molecules in the system. By working within the same rotating and displaced frames as mentioned for the single-molecule scenario, we end up with the following Schrödinger equation
\begin{equation}\label{Eq:App:Schrod:after:Many}
    i \hbar \pdv{\vert\Tilde{\Psi}_{N}(t)\rangle}{t}
        = e \sum_{i=1}^{N_\text{mol}}
        \hat{\boldsymbol{R}}_i(t)\cdot \hat{\boldsymbol{E}}(t)
        \vert\Tilde{\Psi}_N(t)\rangle,
\end{equation}
where we have denoted $\vert\Tilde{\Psi}_N(t)\rangle$ the joint state between the many-molecules and the field. Following the same steps as in the single-molecule analysis, where we neglected the contribution at all times of all continuum and bounded states (different from the ground and first excited ones), we project this equation with respect to $\ket{\psi_{\textbf{m}}} = \ket{\psi_{1,\mathsf{i}}} \otimes \ket{\psi_{2,\mathsf{j}}} \otimes \cdots \otimes \ket{\psi_{N_{\text{mol}},\mathsf{N}}}$, where $\mathsf{i,j},$...$,\mathsf{N} \in \{\mathsf{a},\mathsf{b}\}$. Thus, here $\textbf{m} = \{(1,\mathsf{i}),(2,\mathsf{j}),\cdots,(N_{\text{mol}}, \mathsf{N})\}$ denotes in what state each of the molecules is. By implementing this projection, we get
\begin{equation}
    i \hbar \dv{\ket{\Phi_{\textbf{m}}(t)}}{t}
        = e \sum_{\vb{n}}
        \bigg[
            \bra{\psi_{\vb{m}}}\sum_{i=1}^{N_\text{mol}}
            \hat{\boldsymbol{R}}_i(t)
            \ket{\psi_{\vb{n}}}\cdot \hat{\boldsymbol{E}}(t)
            \ket{\Phi_{\textbf{n}}(t)}
        \bigg],
\end{equation}
where the sum over $\vb{n}$ runs through all the possible combinations of states in the molecules. Let us consider the scenario where it is very unlikely for a single molecule to perform a transition from a bonding to an antibonding state, such that at the end of the HHG process, almost all molecules end up in the initial state except one, which we allow undergoing a bonding-antibonding transition. This means that in the summation over $\vb{n}$ we consider only those elements for which $\vb{n} = \bar{\mathsf{b}} := \{\vb{m}: \mathsf{i}=\mathsf{j}=\cdots=\mathsf{N} = \mathsf{b}\}$ and $\vb{n} = \bar{\mathsf{a}}_k:= \{\vb{m}: \mathsf{i}=\mathsf{j}=\cdots\neq\mathsf{k}\neq \cdots =\mathsf{N} = \mathsf{b} \ \& \ \mathsf{k} = \mathsf{a}\}$, i.e., there is at least one molecule (the $k$th molecule) which is in an antibonding state. Under this assumption, and having in mind that the $\braket{\psi_{\mathsf{a}}}{\psi_{\mathsf{b}}}=0$, our system of equations reads
\begin{align}
     &i \hbar \dv{\ket{\Phi_{\bar{\mathsf{b}}}(t)}}{t}
        = N_{\text{mol}}\boldsymbol{\mu}_{\mathsf{bb}}(t)\cdot \hat{\boldsymbol{E}}(t)\ket{\Phi_{\bar{\mathsf{b}}}(t)}
        + \boldsymbol{\mu}_{\mathsf{ba}}(t)\cdot \hat{\boldsymbol{E}}(t)
            \sum_{k=1}^{N_{\text{mol}}}
            \ket{\Phi_{\bar{\mathsf{a}}_k}(t)},
    \\
    &i \hbar \dv{\ket{\Phi_{\bar{\mathsf{a}}_k}(t)}}{t}
        = \big[
            \boldsymbol{\mu}_{\mathsf{aa}}(t) + (N_{\text{mol}}-1)\boldsymbol{\mu}_{\mathsf{bb}}(t)
        \big]\cdot \hat{\boldsymbol{E}}(t)
        \ket{\Phi_{\bar{\mathsf{a}}_k}(t)}
        + \boldsymbol{\mu}_{\mathsf{ab}}(t)\cdot \hat{\boldsymbol{E}}(t)
            \ket{\Phi_{\bar{\mathsf{b}}}(t)}
        \quad \forall k \in \{1,2,\cdots,N_{\text{mol}}\}.
\end{align}

Note that these equations are very similar to the ones we have solved previously. In fact, once taking into account the initial conditions (all molecules initially in their ground state) and neglecting higher order transition terms as in the molecule scenario, the solution to this differential equation reads

\begin{align}
    &\ket{\Phi_{\bar{\mathsf{b}}}(t)}
        \approx \hat{\mathcal{D}}\big(N_{\text{mol}}\boldsymbol{\chi}_{\mathsf{b}}(t,t_0)\big)
            \ket{\bar{0}},
    \\
    &\ket{\Phi_{\bar{\mathsf{a}_k}}(t)}
        \approx 
        - \dfrac{i}{\hbar}\int^t_{t_0} \dd t_1 
                \hat{\mathcal{D}}\Big(\boldsymbol{\chi}_{\mathsf{a}}(t,t_1)+
                (N_{\text{mol}}-1)\boldsymbol{\chi}_{\mathsf{b}}(t,t_1)\Big)
                \hat{M}_{\mathsf{ab}}(t_1)
                \hat{\mathcal{D}}\big(N_{\text{mol}}\boldsymbol{\chi}_{\mathsf{b}}(t_1,t_0)\big)
                \ket{\bar{0}},\label{Eq:App:antibond:many}
\end{align}

When working with a larger number of molecules, we can approximate $N_{\text{mol}}-1 \approx N_{\text{mol}}$. Furthermore, by further approximating $N_{\text{mol}}\boldsymbol{\chi}_{\mathsf{b}}(t,t_1) + \boldsymbol{\chi}_{\mathsf{a}}(t,t_1) \approx N_{\text{mol}}\boldsymbol{\chi}_{\mathsf{b}}(t,t_1)$, we write the state in Eq.~\eqref{Eq:App:antibond:many} as
\begin{equation}\label{Eq:App:ant:approx:expanded}
    \ket{\Phi_{\bar{\mathsf{a}}}(t)}
        = - 
            \dfrac{i}{\hbar}
            \hat{\mathcal{D}}
                \big(
                     N_{\text{mol}}\boldsymbol{\chi}_{\mathsf{b}}(t,t_0)
                \big) \int^t_{t_0} \dd t_1 
        e^{\theta_{\mathsf{b}}} 
                \boldsymbol{\mu}_{\mathsf{ab}}(t)\cdot
                \big(
                    \hat{\boldsymbol{E}}(t_1) 
                    + \boldsymbol{E}_{\text{cl}}^{(\mathsf{b})}(t_1)
                \big) \ket{\bar{0}},
\end{equation}
where we have defined
\begin{align}
    &\theta_{\mathsf{b}}
        =N_{\text{mol}}^2\sum_q \dfrac12 
            \Big[
                \chi^{(q)}_{\mathsf{b}}(t,t_1)
                \big(
                    \chi_{\mathsf{b}}^{(q)}(t_1,t_0)
                \big)^*
                -
                \big(
                    \chi^{(q)}_{\mathsf{b}}(t,t_1)
                \big)^*
                    \chi^{(q)}_{\mathsf{b}}(t_1,t_0)
            \Big],
 \\
    &
    \boldsymbol{E}^{(\mathsf{b})}_{\text{cl}}(t)
        = -i N_{\text{mol}}\sum_q \vb{g}(\omega_q)
            \Big[
                \big(
                    \chi^{(q)}_{\mathsf{b}}(t,t_0)
                \big)^*e^{i\omega_q t}
                -
                \chi^{(q)}_{\mathsf{b}}(t,t_0)
                    e^{-i\omega_q t}
            \Big].
\end{align}

Then, the joint state of the system is given by
\begin{align}
    \ket{\Tilde{\Psi}(t)}
        &= \dfrac{1}{\sqrt{N}}
            \bigg[
                \ket{\psi_{\bar{\mathsf{b}}}}\ket{\Phi_{\bar{\mathsf{b}}}(t)}
                + \sum^{\text{N}_{\text{mol}}}_{k=1} \ket{\psi_{\bar{\mathsf{a}}_k}}\ket{\Phi_{\bar{\mathsf{a}}_k}}
            \bigg]
        =\dfrac{1}{\sqrt{N}}
            \Big[
                \ket{\psi_{\bar{\mathsf{b}}}}\ket{\Phi_{\bar{\mathsf{b}}}(t)}
                + \sqrt{N_{\text{mol}}} \ket{\psi_{\bar{\mathsf{a}}}}\ket{\Phi_{\bar{\mathsf{a}}}}
            \Big],
\end{align}
where, in going from the first to the second equality, we have taken into account that all the states of the form $\ket{\Phi_{\bar{\mathsf{a}}_k}}$ are independent on $k$. For this reason we remove the index hereupon. Moreover, we defined $\ket{\psi_{\bar{\mathsf{a}}}} = (1/\sqrt{N_{\text{mol}}})\sum_{k=1}^{N_{\text{mol}}}\ket{\psi_{\bar{\mathsf{a}}_k}}$. Note that these results can be obtained from the single-molecular analysis by changing $\boldsymbol{\mu}_{\mathsf{bb}}(t) \to N_{\text{mol}}\boldsymbol{\mu}_{\mathsf{bb}}(t)$ and $\boldsymbol{\mu}_{\mathsf{ij}}(t) \to \sqrt{N_{\text{mol}}}\boldsymbol{\mu}_{\mathsf{ij}}(t)$. Moreover, note that in the case that $N_{\text{mol}}$ becomes extremely large, it can dominate over the bonding-bonding contribution. We expect that, in this regime, one has to go further than the first-order perturbation theory under which we have been working. Thus, in the numerical calculations we restrict to situations where the probability of having (many-molecule) events ending up in an antibonding state are smaller than those where all molecules end up in a bonding state.

These are the expressions used in our numerical calculations. However, in the remnant of this Appendix material, we will also introduce the following notation
\begin{equation}\label{Eq:App:ant:h:functions}
    \ket{\Phi_{\bar{\mathsf{a}}}(t)}
        = \hat{\mathcal{D}}
            \big(
                \boldsymbol{\chi}_{\mathsf{b}}(t,t_0)
            \big)
            \sum_{q=1}
            \Big[
                h_1^{(q)}(t)
                    \ket{1_q}\bigotimes_{q'\neq q}\ket{0}
                + h_2^{(q)}(t)\ket{\bar{0}}
            \Big],
\end{equation}
where we have defined
\begin{align}
    & h_1^{(q)}
        = -\dfrac{i}{\hbar}
            \sqrt{N_{\text{mol}}}
            \int^t_{t_0} \dd t_1
                e^{\bar{\theta}_{\mathsf{b}}}
                \vb{g}(\omega_q)
                \cdot \boldsymbol{\mu}_{\mathsf{ab}}(t_1)e^{i\omega_q t_1}
        \label{Eq:App:h1},
    \\
    &h_2^{(q)}(t)
        = -\dfrac{i}{\hbar}
            \sqrt{N_{\text{mol}}}
            \int^t_{t_0} \dd t_1
                e^{\bar{\theta}_{\mathsf{b}}}
                \boldsymbol{\mu}_{\mathsf{ab}}(t_1)
                \cdot
                \boldsymbol{E}^{(q)}_{\text{cl}}(t_1)
                \label{Eq:App:h2}.
\end{align}

At some point in the rest of the Appendices, we shall work with $H_2(t) \equiv \sum_q h_2^{(q)}(t)$.

\section{Computing and lower bounding different entanglement measures and witnesses}\label{App:Final}

    \subsection{Characterization of the light-matter entanglement}\label{App:light:matter}
    In this subsection, we show how we computed the entropy of entanglement for characterizing the light-matter entanglement, ultimately leading to the results shown in Fig.~\eqref{Fig:light:elec:ent} of the main text. We first trace either over the electronic or the field degrees of freedom. In our case, we selected the first approach, since effectively the electron can be studied as a two-level system where only the bonding and antibonding states are populated. For the second approach, when working with the quantum optical degrees of freedom instead, because of the presence of the displacement operators, a continuous set of modes should be considered.

    Thus, the electronic state, once the quantum optical modes are traced out, is given by
    \begin{equation}
        \begin{aligned}\label{Eq:App:rho:elec}
            \hat{\rho}_{\text{elec}}(t)
                &= \tr_f\big(
                        \lvert\Tilde{\Psi}(t)\rangle 
                        \!\langle \Tilde{\Psi}(t)\rvert
                    \big)
                \\
                &=\dfrac{1}{\mathcal{N}}
                    \Big[
                        \dyad{\psi_{\mathsf{b}}}
                        + \mathcal{N}_{\mathsf{a}}
                            \dyad{\psi_{\mathsf{a}}}
                        + \sqrt{\mathcal{N}_{\mathsf{a}}}
                         \braket{\bar{\Phi}_{\mathsf{a}}(t)}{\Phi_{\mathsf{b}}(t)} 
                         \dyad{\psi_{\mathsf{b}}}{\psi_{\mathsf{a}}}
                        +  \sqrt{\mathcal{N}_{\mathsf{a}}}
                         \braket{\Phi_{\mathsf{b}}(t)}{\bar{\Phi}_{\mathsf{a}}(t)}
                         \dyad{\psi_{\mathsf{a}}}{\psi_{\mathsf{b}}}
                    \Big],
        \end{aligned}
    \end{equation}
    where we have that
    \begin{equation}
        \begin{aligned}
            \sqrt{\mathcal{N}_a}\braket{\Phi_{\mathsf{b}}(t)}{\bar{\Phi}_{\mathsf{a}}(t)}
                =
                    H_2(t)
        \end{aligned}
    \end{equation}
    with the $H_2(t)$ and $h_1^{(q)}(t)$ functions defined in Eqs.~\eqref{Eq:App:h1} and \eqref{Eq:App:h2}. The density matrix shown in Eq.~\eqref{Eq:App:rho:elec} has the form of a single-qubit matrix which one could easily numerically diagonalize. We did this using the \texttt{Scipy} package of Python \cite{2020SciPy-NMeth}, to find two eigenvalues $\{\lambda_1, \lambda_2\}$. With this, the entropy of entanglement can be easily computed as
    \begin{equation}
        S\big(\hat{\rho}_{\text{elec}}(t)\big)
            = - \tr
                \big[
                    \hat{\rho}_{\text{elec}}(t)
                    \log_2 \hat{\rho}_{\text{elec}}(t)
                \big]
            = - \sum_{i=1}^2 \lambda_i \log_2 \lambda_i.
    \end{equation}
    \subsection{Characterization of the entanglement between the harmonic modes}\label{App:light:light:ent}
    In this subsection, we show how the entanglement between the harmonic modes for the different cases studied in the main text is. Specifically, we present the lower bounds used for obtaining the results shown in Fig.~\ref{Fig:light:light:ent:cond}~(c) and Fig.~\ref{Fig:light:light:ent:LN}. For the sake of clarity, we present each of the cases separately.
    
    \subsubsection{Entanglement between the harmonic modes when conditioning the electron to be in an antibonding state}
    In this case, we consider the harmonic modes to be divided into two sets, $A$ and $B$, such that,
    when the electron is conditioned to be found in an antibonding state, i.e., when projecting Eq.~\eqref{Eq:App:ant:h:functions} with respect to $\ket{\psi_{\mathsf{a}}}$, the quantum optical state can be written as
    \begin{equation}\label{Eq:App:Ant:light:light:ent}
        \begin{aligned}
        \ket{\Phi_{\mathsf{a}}(t)}
            = \dfrac{\hat{\mathcal{D}}\big(\boldsymbol{\chi}_{\mathsf{b}}(t,t_0)\big)}{\sqrt{\mathcal{N}_{\mathsf{a}}}}
                \bigg[
                    \Big(
                        \sum_{q\in A}h_1^{(q)}(t)\ket{1_q}\ket{0_{q'\neq q}}
                    \Big)\ket{\bar{0}_B}
                    + \ket{\bar{0}_A}
                    \Big(
                        \sum_{q\in B}h_1^{(q)}(t)\ket{1_q}\ket{0_{q'\neq q}}
                    \Big)
                    +H_2(t)\ket{\bar{0}_A}\ket{\bar{0}_B}
                \bigg],
        \end{aligned}
    \end{equation}
    where we have defined $\ket{\bar{0}_A}=\bigotimes_{q\in A}\ket{0_q}$ (same holds for $B$). In order to characterize the amount of entanglement in this state by using the entropy of entanglement, it is more convenient for us to work under a different basis set. Specifically, we express each of the subsystems in terms of the basis set spanned by the following states (we use the set $A$ as an example)
    \begin{equation}\label{Eq:App:basis}
        \Big\{
            \ket{\Tilde{0}_A} 
                = \hat{\mathcal{D}}
                    \big(\boldsymbol{\chi}_{\mathsf{b}}(t,t_0)\big)
                        \ket{\bar{0}_A},
            \ket{\Tilde{1}_A}
                = \dfrac{1}{\sqrt{\mathcal{N}_A}}
                    \sum_{q\in A}
                        h_q^{(q)}(t)
                        \hat{\mathcal{D}}
                        \big(
                            \boldsymbol{\chi}_{\mathsf{b}}(t,t_0)
                        \big)
                        \ket{1_q}\ket{0_{q'\neq q}}
            , \cdots
        \Big\},
    \end{equation}
    where the ellipsis, i.e. the ``$\cdots$'', represents orthonormal states to the other two. We note that the two states we have considered contain, within a displaced frame, either zero or one excitation of the field modes. Thus, the extra orthonormal states included in the ellipsis can be obtained, for instance, by means of the Gram-Schmidt decomposition and using Fock states (within the displaced frame) containing more than two excitations. Nevertheless, in our case it is enough to just consider the ones explicitly shown in Eq.~\eqref{Eq:App:basis} as, by means of these, we can rewrite Eq.~\eqref{Eq:App:Ant:light:light:ent} as
    \begin{equation}
        \ket{\Phi_{\mathsf{a}}(t)}
            = \dfrac{1}{\sqrt{\mathcal{N}_{\mathsf{a}}}}
                \Big[
                    \sqrt{N_A}\ket{\Tilde{1}_A}\ket{\Tilde{0}_B}
                    +\sqrt{N_B}\ket{\Tilde{0}_A}\ket{\Tilde{1}_B}
                    + H_2(t) \ket{\Tilde{0}_A}\ket{\Tilde{0}_B}
                \Big].
    \end{equation}

    After tracing out one of the subsystems' degrees of freedom (for instance $B$), we obtain the following reduced density matrix for the other state
    \begin{equation}
        \begin{aligned}
        \hat{\rho}^A_{\mathsf{a}}(t)
            &= \tr_B\big(\dyad{\Phi_{\mathsf{a}}(t)})
            \\
            &=\dfrac{1}{N_{\mathsf{a}}}
                \Big[
                    N_A \dyad{\Tilde{1}_A}
                    + (N_B + \abs{H_2(t)}^2)\dyad{\Tilde{0}_A}
                    +H_2^*(t) \sqrt{N_A}
                        \dyad{\Tilde{1}_A}{\Tilde{0}_A}
                    + H_2(t)\sqrt{N}_A
                        \dyad{\Tilde{0}_A}{\Tilde{1}_A}
                \Big],
        \end{aligned}
    \end{equation}
    for which the entropy of entanglement can be computed following a similar procedure to that of Appendix~\ref{App:light:matter}.

    \subsubsection{Entanglement between the harmonic modes when conditioning the electron to be in a \emph{localized} right or left state}
    The calculation of the entropy of entanglement in this case can be obtained in a very straightforward way by redefining $H_2(t) \to 1 + H_2(t)$ in the previous subsection, arising from the contribution of the bonding component not appearing before.
    
\subsubsection{Entanglement between the harmonic modes when the final state of the electron is unknown}
In this subsection, we focus on the case where we have no knowledge about what state has the electron recombined with. For convenience, we work with the bonding and antibonding quantum optical components, such that after tracing out the electronic degrees of freedom we have for the quantum optical state
\begin{equation}\label{Eq:App:QO:mixed}
    \hat{\rho}_f(t)
        = \dfrac{1}{\mathcal{N}}
            \big[
                \dyad{\Phi_{\mathsf{b}}(t)}
                + \dyad{\Phi_{\mathsf{a}}(t)}
            \big]
        = \dfrac{1}{\mathcal{N}}
            \big[
                \hat{\rho}_{\mathsf{b}}(t)
                + \mathcal{N}_{\mathsf{a}}
                    \hat{\rho}_{\mathsf{a}}(t)
            \big],
\end{equation}
which indeed coincides with Eq.~\eqref{Eq:HHG:QO:Full} of the main text. 

Our aim is to study the amount of entanglement present in the state, Eq.~\eqref{Eq:App:QO:mixed}, between a single mode $\Tilde{q}$ and the rest. However, unlike the previous case, here we are working with mixed states for which the entropy of entanglement is not a valid entanglement measure. Instead, we work with the logarithmic negativity \cite{amico_entanglement_2008}, defined as $E_N(\hat{\rho}):=\log_2\big(2 \mathsf{N} + 1)$ where $\mathsf{N}$ is the negativity, i.e., the sum of all negative eigenvalues (in absolute value) of the partial transpose of $\hat{\rho}$ with respect to one of the subsystems. In our case, if we define subsystems $\bar{A} := \{\Tilde{q}\}$ and $\bar{B}:= \{q: \forall q \neq \Tilde{q}\}$, and consider the partial transpose with respect to subsystem $\bar{B}$, we then have
\begin{equation}\label{Eq:App:PT:rho}
    \hat{\rho}^{T_{\bar{B}}}_f(t)
        = \dfrac{1}{\mathcal{N}}
            \Big[
                \hat{\rho}_{\mathsf{b}}(t)
                + \mathcal{N}_{\mathsf{a}}
                \hat{\rho}^{T_B}_{\mathsf{a}}(t)
            \Big],
\end{equation}
where we have taken into account that $\hat{\rho}_{\mathsf{b}}(t)$ is actually a pure separable state and does not get affected by the partial transpose operation.

In order to compute the negativity, we first need to find the negative eigenvalues of Eq.~\eqref{Eq:App:PT:rho}. The main problem here is that, because of the different displacement operations appearing in the definitions of $\hat{\rho}_{\mathsf{a}}(t)$ and $\hat{\rho}_{\mathsf{b}}(t)$, we cannot look for a proper basis as in Eq.~\eqref{Eq:App:basis} which allows us to make the calculations manageable. Therefore, in order to \emph{formally} compute the negativity, we would have to consider the full basis set which, in Fock representation, is composed of hundreds of states. Thus, instead of computing the logarithmic negativity exactly, we propose a lower bound easier to handle numerically. For this, we first take into account that, given three Hermitian matrices $\mathcal{A},\mathcal{B}$ and $\mathcal{C} = \mathcal{A} + \mathcal{B}$, with respective eigenvalues $\{a_1 > a_2 > \cdots > a_n\}$, $\{b_1 > b_2 > \cdots > b_n\}$ and $\{c_1 > c_2 > \cdots > c_n\}$, the following relationship holds for their eigenvalues \cite{weyl_asymptotische_1912,fulton_eigenvalues_2000}
\begin{equation}
    c_{i+j-1} \leq a_i + b_j, \quad \text{for } i+j-1\leq n.
\end{equation}

If we focus on the potential negative eigenvalues that $\mathcal{C}$ could have, we can write for their absolute value
\begin{equation}
    \abs{c_{i+j-1}} \geq \abs{a_i + b_j},
\end{equation}
such that if we identify $\mathcal{A}=\hat{\rho}_{\mathsf{a}}^{T_{\bar{B}}}$, $\mathcal{B} = \hat{\rho}_{\mathsf{b}}(t)$ and $\mathcal{C}= \hat{\rho}^{T_{\bar{B}}}_{f}(t)$, we can write for the negativity of $\mathcal{C}$
\begin{equation}
    \mathsf{N}
        = \sum_{c_i < 0} \abs{c_i}
        \geq \sum_{a_i + b_j < 0}
        \abs{\min_{i,j}(a_i + b_j)}
        \geq \abs{\min_i a_i},
\end{equation}
where in the last inequality we have taken into account that $\mathcal{B}$ does not have, by definition, negative eigenvalues, and its lowest eigenvalue is zero since it is a pure separable state. Thus, the last inequality corresponds to the minimum eigenvalue found for $\mathcal{A}$. Having this relationship in mind, we propose the following lower bound for the logarithmic negativity
\begin{equation}
    E_N(\hat{\rho}) \geq \mathcal{E}_N(\hat{\rho})
    = \log_2
        \Big(
            2 \abs{\min_i a_i} + 1
        \Big),
\end{equation}
which is what is actually shown in Fig.~\ref{Fig:light:light:ent:LN} of the main text.
\end{document}